\documentclass[sigconf, nonacm]{acmart}

\newcommand\vldbdoi{XX.XX/XXX.XX}
\newcommand\vldbpages{XXX-XXX}
\newcommand\vldbvolume{14}
\newcommand\vldbissue{1}
\newcommand\vldbyear{2020}
\newcommand\vldbauthors{\authors}
\newcommand\vldbtitle{\shorttitle} 
\newcommand\vldbavailabilityurl{https://github.com/patgiri/KmerCo-Main}
\newcommand\vldbpagestyle{plain}

\usepackage{amsmath}
\usepackage{hyperref}
\usepackage{url}
\usepackage{algorithm}
\usepackage{algpseudocode}
\usepackage{caption}
\usepackage{subcaption}
\usepackage{pgfplots}
\usepackage{multirow}
\pgfplotsset{compat=1.6}
\usepackage{textcomp}
\usepackage{balance}
\usepackage{xcolor}
\usetikzlibrary{patterns}
\usepackage{tikz}
\usetikzlibrary{backgrounds,automata}
\usepackage{float}
\usepackage{ulem}

\algblock{Input}{EndInput}
\algnotext{EndInput}
\algblock{Output}{EndOutput}
\algnotext{EndOutput}

\makeatletter
\newcommand\resetstackedplots{
    \makeatletter
    \pgfplots@stacked@isfirstplottrue
    \makeatother
}
\makeatother

\definecolor{mygray}{gray}{0.6}
\definecolor{mygreen}{rgb}{0.0, 0.5, 0.69}
\definecolor{myblue}{rgb}{0.2, 0.2, 0.6}
\definecolor{myred}{rgb}{0.8, 0.25, 0.33}
\definecolor{ao}{rgb}{0.0, 0.5, 0.0}
\definecolor{cadmiumgreen}{rgb}{0.0, 0.42, 0.24}

\begin{document}
\title{KmerCo: A lightweight K-mer counting technique with a tiny memory footprint}  

\author{Sabuzima Nayak}
\orcid{0000-0001-7572-1054}
\affiliation{%
  \institution{National Institute of Technology Silchar}
  \city{Silchar}
  \state{Assam}
  \country{India}
  \postcode{788010}
}
\email{sabuzima\_rs@cse.nits.ac.in}

\author{Ripon Patgiri}
\orcid{0000-0002-9899-9152}
\affiliation{%
  \institution{National Institute of Technology Silchar}
  \city{Silchar}
  \state{Assam}
  \country{India}
  \postcode{788010}
}
\email{ripon@cse.nits.ac.in}


\begin{abstract}
K-mer counting is a requisite process for DNA assembly because it speeds up its overall process. The frequency of K-mers is used for estimating the parameters of DNA assembly, error correction, etc. The process also provides a list of district K-mers which assist in searching large databases and reducing the size of de Bruijn graphs. Nonetheless, K-mer counting is a data and compute-intensive process. Hence, it is crucial to implement a lightweight data structure that occupies low memory but does fast processing of K-mers. We proposed a lightweight K-mer counting technique, called KmerCo that implements a potent counting Bloom Filter variant, called countBF. KmerCo has two phases: insertion and classification. The insertion phase inserts all K-mers into countBF and determines distinct K-mers. The classification phase is responsible for the classification of distinct K-mers into trustworthy and erroneous K-mers based on a user-provided threshold value. We also proposed a novel benchmark performance metric. We used the Hadoop MapReduce program to determine the frequency of K-mers. We have conducted rigorous experiments to prove the dominion of KmerCo compared to state-of-the-art K-mer counting techniques. The experiments are conducted using DNA sequences of four organisms. The datasets are pruned to generate four different size datasets. KmerCo is compared with Squeakr, BFCounter, and Jellyfish. KmerCo took the lowest memory, highest number of insertions per second, and a positive trustworthy rate as compared with the three above-mentioned methods.  
\end{abstract}
\maketitle

\pagestyle{\vldbpagestyle}
\begingroup\small\noindent\raggedright\textbf{PVLDB Reference Format:}\\
\vldbauthors. \vldbtitle. PVLDB, \vldbvolume(\vldbissue): \vldbpages, \vldbyear.\\
\href{https://doi.org/\vldbdoi}{doi:\vldbdoi}
\endgroup
\begingroup
\renewcommand\thefootnote{}\footnote{\noindent
This work is licensed under the Creative Commons BY-NC-ND 4.0 International License. Visit \url{https://creativecommons.org/licenses/by-nc-nd/4.0/} to view a copy of this license. For any use beyond those covered by this license, obtain permission by emailing \href{mailto:info@vldb.org}{info@vldb.org}. Copyright is held by the owner/author(s). Publication rights licensed to the VLDB Endowment. \\
\raggedright Proceedings of the VLDB Endowment, Vol. \vldbvolume, No. \vldbissue\ %
ISSN 2150-8097. \\
\href{https://doi.org/\vldbdoi}{doi:\vldbdoi} \\
}\addtocounter{footnote}{-1}\endgroup

\ifdefempty{\vldbavailabilityurl}{}{
\vspace{.3cm}
\begingroup\small\noindent\raggedright\textbf{PVLDB Artifact Availability:}\\
The source code, data, and/or other artifacts have been made available at \url{\vldbavailabilityurl}.
\endgroup
}
\sloppy
\section{Introduction}
Gregor Mendel discovered the genes in peas \cite{peas} whereas rules of genes were discovered in red bread mold \cite{mold}. DNA was discovered in salmon \cite{salmon} and some information regarding the encapsulation of DNA was known from tardigrades \cite{Tardigrades}. Chromosomes were first noticed in mealworms, likewise, sex chromosomes were discovered in beetles \cite{mealworm} whereas its function and replication were explored in platypus and fish \cite{Platypus}. This illuminates the importance of DNA sequencing of organisms. Genome sequencing enhances our understanding regarding the complexity of the evolution of life, its functioning, and the protection of our biodiversity. The DNA sequencing of all organisms is essential because it helps to compare the DNA sequences and some unique features of the organisms. Although this may be important, DNA sequencing is a complex process. The next generation sequencing is efficiently generating genomic data by reading short reads. A $read$ is a DNA subsequence of length 20-30000 bases. The $read$s have many overlapping regions with other $read$s. Moreover, errors are introduced in the $read$s during the electrical and chemical processing. DNA assembler removes errors and arranges the $read$s to obtain the DNA sequence. Hence, the foremost tasks of DNA assembler are error removal and identification of distinct K-mers. A single process, i.e., K-mer counting completes both tasks. 

\subsection{K-mer counting}
\begin{figure}[!ht]
    \centering
    \includegraphics[width=0.9\columnwidth]{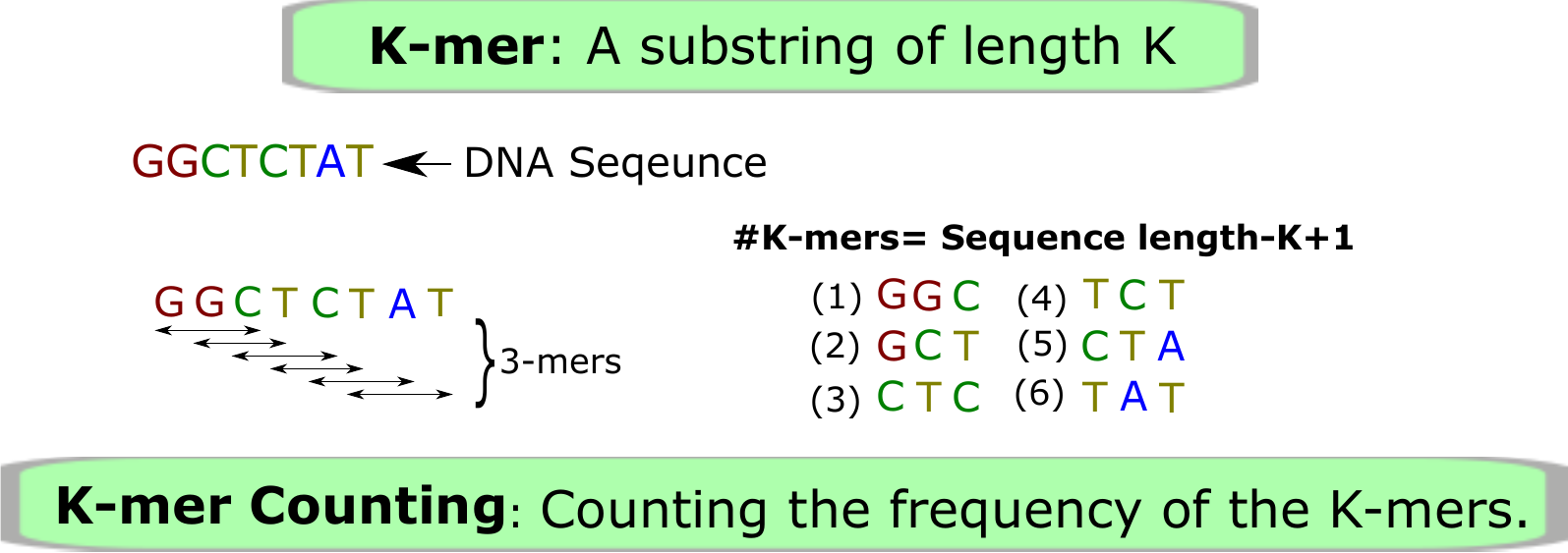}
    \caption{Graphical explanation of K-mer.}
    \label{kmer}
\end{figure}

The “mer” is a Greek word that means “part”. K-mer means a sub-string of length K. K-mers of a DNA sequence are all possible consecutive K-mers of length K. Figure \ref{kmer} illuminates an example for better understanding. Suppose \textbf{GGCTCTAT} is a DNA sequence. The left side of the figure represents the method to consider the consecutive 3-mers and the right side of the figure lists the 3-mers. The number of K-mers in a DNA sequence is sequence length-K+1 where sequence length is the number of nucleotide present in the DNA sequence. K-mer counting is a process of counting the frequency of the K-mers in a DNA sequence \cite{Assembly}. This process takes DNA files as input, extracts K-mers, and counts their frequency. As output, it generates distinct K-mers and acts as a classifier to eliminate the low-frequency K-mers. 

\textbf{Why do we count the K-mers?} The answer is as follows- (a) Speedup DNA Assembly: Some DNA assembly techniques speed up the overall process using K-mer counting. For instance, overlap layout consensus (OLC) searches the read overlaps which is a slow process. K-mer counting speeds up this process \cite{MSPKmer}. 
(b) Calculating DNA Assembly parameters: The count of distinct and trustworthy K-mers helps in determining the parameters required in the DNA Assembly process. 
(c) Error correction: K-mers are highly repetitive even in a small fragment of a DNA sequence \cite{FLAMM}. For this reason, a few times occurring K-mer is an error. An error occurs during genomic data collection due to incorrect reading of a few bases, or the addition or removal of a DNA fragment. 
(d) Metagenomics: It identifies the K-mers present in the DNA sequence \cite{metagenomic}, for example, verifying the presence of a protein in a sample DNA sequence.
(e) Searching in large datasets: The distinct K-mers generated by the K-mer counting techniques are used to search in a DNA bank to identify the DNA sequence having those K-mers. 
(f) Small size de Bruijn graph: K-mer counting provides the distinct K-mers and their frequency which helps in quick construction and size reduction of the de Bruijn graph. 

\subsection{Bloom Filter}
\begin{figure}[!ht]
    \centering
    \includegraphics[width=0.9\columnwidth]{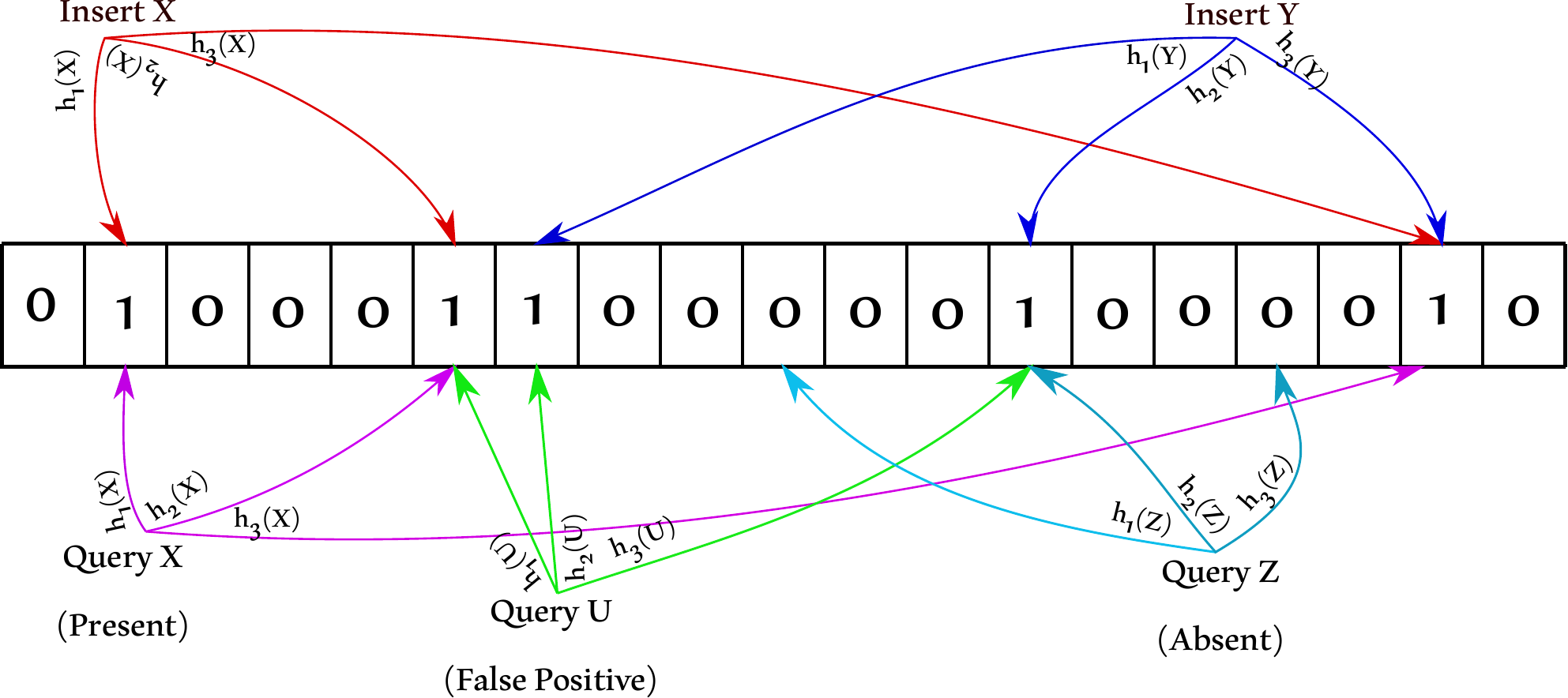}
    \caption{\textbf{Architecture of Standard Bloom Filter using three hash functions.}}
    \label{bf_arch}
\end{figure}

Bloom Filter \cite{Bloom,BF1} is a probabilistic simple bit array data structure used for determining the membership of an item. Bloom Filter does not store the original data; rather the input item is mapped to Bloom Filter bits which helps to store many items using a small-sized bit array. Figure \ref{bf_arch} illustrates the architecture and operation of the standard Bloom Filter. Bloom Filter is a bit array where each bit is set to either 0 or 1. Initially, all slots are set to 0. Bloom Filter performs two operations: insertion and query. An input item is hashed by the hash function(s), say $k_h$. The hashed value determines the slot location which is set to 1. The $k_h$ slots are set to 1, as illustrated by inserting items X and Y in Figure \ref{bf_arch}. The query operation follows the same procedure as the insertion operation to obtain the bit locations. If all slots are 1, then the item is present; if at least one slot is 0, then the item is absent. In Figure \ref{bf_arch}, item X is present whereas item Z is absent. The time complexity of insertion and query operation is $O(k_h)\approx O(1)$. Consider the query of item U in the figure, U is not inserted but during the query operation, all slots are 1. This situation is created by the insertion of X and Y. The slots obtained by the hashed values of U are colliding with the slots of X and Y. The true response returned by Bloom Filter in such a query operation is called a false positive. Therefore, the main aim while proposing a new Bloom Filter variant is to reduce the false positive probability (FPP). A variant of Bloom Filter is Counting Bloom Filter (CBF) \cite{CBF} which is proposed to reduce FPP. Each slot is partitioned into a bit and a counter of a few bits. Initially, all slots are set to 0. It follows the same procedure to obtain the slots, the bit and counter are set to 1. Only the counter is incremented if a new item is hashed to the same slot. The counter keeps the frequency of the items. However, CBF has a counter overflow issue.

\subsection{Challenges}
K-mer counting is a data-intensive and compute-intensive task. K-mer counting takes half of the total computation time in DNA assembly techniques \cite{Chikhi}. It is data-intensive because a genomic file has millions or billions of K-mers. Each K-mer needs to be processed to verify whether it is encountered for the first time to include in the list of distinct K-mers; otherwise, increment the frequency of the K-mer. The processing of such a huge volume of data is compute-intensive. The hashtable-based techniques such as Jellyfish \cite{jellyfish} require a high memory footprint. Moreover, disk-based techniques such as KMC2 \cite{KMC2} are inefficient as it takes huge time to process as compared to lightweight ones. There is a requirement for a data structure that is faster, has a low memory footprint, and is efficient. One such data structure is Bloom Filter which is a solution to all these issues.  

There are many Bloom Filter-based K-mer counting techniques. However, the Bloom Filter is not responsible for counting K-mers. The Bloom Filter is used for membership checking or filtering of the first encounter K-mers. The techniques maintain a hashtable to keep the count of the K-mers. Hence, the techniques require more memory because they maintain two data structures. However, CBF and its variants can be implemented to keep the count of the K-mers. These Bloom Filters do not provide exact K-mer count; however, the main focus of the K-mer counting technique is the identification of distinct K-mers, and the classification of K-mers into trustworthy and erroneous K-mers rather than the exact count of K-mers. The count of K-mer merely helps in classification. Thus, a CBF or its variant is more efficient for a K-mer counting technique because a single data structure is capable of both storing K-mers and keeping the count of K-mers. 

There is a lack of experimental benchmark that evaluates the performance of the K-mer counting techniques. The state-of-the-art research articles depict the experimental results in tabular form that does not adduce performance. The listing of distinct, trustworthy, and erroneous K-mers in a table does not convey any information regarding the deviation of the presented techniques from the correct values. Notably, only the RAM usage and insertion time are the measurements of performance. Regardless, accuracy is an important performance metric that is neglected due to the lack of an experimental benchmark. 

\subsection{Contributions}
We proposed a fast, efficient, and lightweight K-mer counting technique, called KmerCo, which implements a fast CBF variant called countBF \cite{countBF}. Furthermore, we have proposed a novel benchmark performance metric for the K-mer counting technique. KmerCo quickly processes the K-mers while maintaining a low memory footprint. It processes millions of K-mers within a few seconds. KmerCo classifies the K-mers based on the user input threshold value. It provides countBF and three files, i.e., distinct, trustworthy, and erroneous K-mers as output where countBF can be used for querying K-mers and their frequency. The distinct file contains the list of all distinct K-mers present in the input DNA file. The trustworthy file contains the list of all K-mers having a frequency more than the user input threshold value. The erroneous file contains the list of all K-mers having a frequency less than or equal to the user input threshold value. 

We have conducted extensive experiments on KmerCo using four real datasets of different organisms to measure its various performance parameters. We have trimmed the datasets to have different-sized datasets to observe the change in KmerCo performance with the change in dataset size. We have considered two different K lengths: 28 and 55 to notice the efficiency of KmerCo with different K-mer lengths. We have compared KmerCo with three K-mer counting techniques: Squeakr (a Bloom Filter-based technique), BFCounter (implements both Bloom Filter and hashtable), and Jellyfish (a hashtable-based technique). Our proposed K-mer counting benchmark performance metric is the counting of the K-mers using the Hadoop MapReduce program. The Hadoop provides zero error K-mer frequency counts and a list of distinct and trustworthy K-mers. These values help to determine the deviation of distinct and trustworthy K-mers generated by KmerCo and other state-of-the-art techniques. The performance of KmerCo was compared with other techniques based on data structure memory size, insertion time, number of insertions, inserted-to-ignored K-mer ratio, number of insertions per second, and trustworthy rate. KmerCo requires the lowest memory which is $7.08\times$, $115.25\times$, and $8889.08\times$ less memory compared to Squeakr, BFCounter, and Jellyfish, respectively, for the 28-mers Balaenoptera dataset. KmerCo took the highest insertion time in the case of 55-mers because it inserted all K-mers whereas other techniques inserted lesser K-mers. It is important to notice that KmerCo has a zero inserted-to-ignored ratio whereas other techniques have a non-zero ratio with a negative ratio in a few cases. Moreover, KmerCo is the second highest number of insertions per second after Jellyfish which is due to the Jellyfish’s lowest insertion time. Notable, Jellyfish requires the highest memory footprint, i.e., it consumes a minimum of 2368 MB memory footprint in our experiment. Apart from all, KmerCo has a positive trustworthy rate whereas others have a negative one, which indicates that other techniques are classifying many trustworthy K-mers as erroneous. 

Finally, summarising the contributions of this paper as follows- 
\begin{itemize}
    \item Proposed technique, KmerCo, is a fast, efficacious, and lightweight K-mer counting method. 
\item KmerCo implements a counting Bloom Filter which has a low memory footprint and false positive probability, called countBF.
\item KmerCo provides a list of distinct K-mers using countBF.
\item KmerCo classifies the K-mers based on a user-provided threshold value.
\item KmerCo gives countBF and three files, i.e,  distinct, trustworthy, and erroneous K-mers as output.
\item Proposed a novel benchmark performance metric for K-mer counting techniques.
\item KmerCo has high performance compared to other state-of-the-art K-mer counting techniques.
\item KmerCo requires 7.08, 115.25, 8889.08 times less memory compared to Squeakr, BFCounter, and Jellyfish for insertion of 28-mers Balaenoptera dataset. KmerCo has a zero inserted-to-ignored K-mer ratio. Moreover, KmerCo has a positive trustworthy rate in all datasets whereas other techniques have a negative trustworthy rate in the majority of datasets.  
\end{itemize}

\textbf{Why do we use Bloom Filter-based technique instead of a Hadoop-based technique for K-mer counting?}\\
The answer is as follows:
\begin{itemize}
    \item Not an in-memory program: Hadoop program is a heavyweight program that requires many resources such as CPUs, memory, HDD, etc. Contrary, Bloom Filter is implemented for its low memory footprint. It is a lightweight data structure. Therefore, Bloom Filter is a suitable technique for K-mer counting. 
    \item Processing overhead: Hadoop requires network communication between the $Map$ tasks and $Reduce$ tasks which require extra time to complete the entire process due to the network latency. We know that Hadoop MapReduce is a distributed computing platform, and therefore, it requires many computing resources in cluster mode. 
\end{itemize}


\section{Background}
\subsection{countBF}
\begin{figure}[!ht]
    \centering
    \includegraphics[width=0.9\columnwidth]{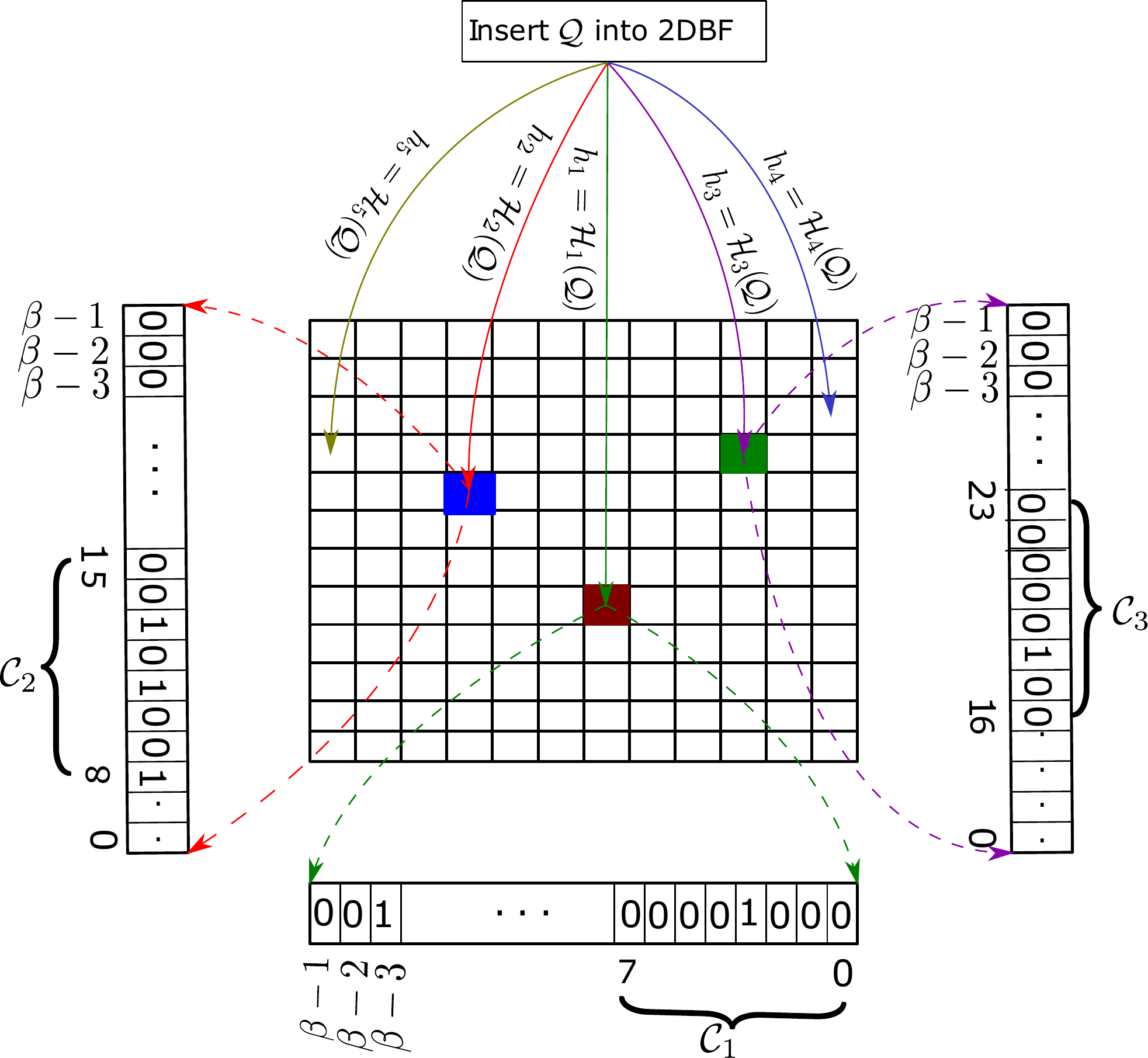}
    \caption{Architecture of countBF with 8-bit counters}
    \label{arc}
\end{figure}

The countBF \cite{countBF} is a CBF variant based on 2DBF \cite{rDBF} which is a two-dimensional integer array. Each slot of countBF is $\beta$-bit length which is partitioned into $\eta$ number of counters of $\alpha$-bit length. The user can define the counter length. Figure \ref{arc} presents the architecture of countBF with an 8-bit counter. As shown in figure, $C_1$, $C_2$, $C_3$, etc are counters. Based on the counter length some bits are unused. The countBF performs two operations: insertion and query. Algorithm \ref{Algo3} depicts the insertion operation where K-mer is the input item. Initially, all slots were set to 0. Let $\mathcal{Q}$ be an item inserted into countBF as presented in Figure \ref{arc}. The $\mathcal{Q}$ is hashed by $k_h$ hash functions. Line 8 of the Algorithm \ref{Algo3} is used to obtain the slot and counter location. The countBF has two predefined masks: extract mask and reset mask. The extract mask extracts corresponding counter values using bit operations. The reset mask helps to reset the corresponding counter value to zero. Let $\mathcal{M}^e_l$ be the extract mask and $\mathcal{M}^r_l$ be the reset mask where $e$ indicates the extract mask, $r$ indicates the reset mask and $l$ is the counter number. The extract mask for the $1^{st}$ counter for an 8-bit counter ($\mathcal{M}^e_1$) is $0x00000000000000FF$ and the reset mask for the $1^{st}$ counter ($\mathcal{M}^r_1$) is $0xFFFFFFFFFFFFFF00$. In Line 9 of the Algorithm \ref{Algo3}, the AND operation is performed between the corresponding slot and the corresponding extract mask to obtain the corresponding counter value with rest bits zero. The counter value is right-shifted to generate only the corresponding counter value. Then the counter value is incremented and to avoid overflow of value to the adjacent counter the new counter value is checked for maximum value. In case, the counter value reaches the maximum value the insert operation is terminated. To avoid the overflow issue applications having high frequency should consider longer counters in countBF. Line 16 of the Algorithm \ref{Algo3} indicates the left shifting of  the new counter value to the required location. Then, an AND operation between the corresponding slot and reset mask removes the old counter value. The new counter value is reflected in the slot by performing OR operation between the slot and the new counter value. This procedure is followed for $k_h$ times. Algorithm \ref{Algo5} outlines the countBF query operation which returns the frequency of the queried item. The counter value is obtained following a similar procedure as in the insertion operation. If the counter value is zero, then the item is absent; otherwise, return the minimum counter value among the $k_h$ slots. 

Let $X$ and $Y$ be countBF dimensions which are prime numbers, $n$ be the total number of input items, and FPP is the false positive probability.\\
The standard Bloom Filter size ($m$)=$\frac{-n log(FPP)}{(log 2)^2}$\\
However, $m$ is a large size for countBF \\
$\therefore$ $v=\sqrt{\frac{m}{128}}$ \\
$X$ is the closest prime number more than $v$\\
$Y$ is the third consecutive prime number from $X$ if all prime numbers are written sequentially.\\
Thus, \begin{equation}
    m_{countBF}=X\times Y\times \beta~~ bits
    \label{eqm}
\end{equation}  \\
The countBF size depends on the number of input items, i.e., $n$.

\subsection{Reverse complement}
The complement of a DNA sequence is obtained by replacing each nucleotide with its complement nucleotide. The nucleotides A, C, G, and T are replaced with T, G, C, and A, respectively. The DNA sequence has another symbol, i.e, N. The N symbol is used to indicate any nucleotide but not a gap. The N remains the same in the complement sequence. The complement sequence is a 3’ to 5’ representation, but DNA sequences are represented from 5’ to 3’. Henceforth, the complement sequence is reversed which is called the reverse complement of the forward/original DNA sequence \cite{RC}. Example: consider GGCTCTAT as the original DNA sequence, its complement sequence is CCGAGATA and the reverse complement is ATAGAGCC. 

\subsection{Canonical K-mer}
It is the lexicographical smaller K-mer between the original ($K\text{-}mer$) and the reverse complement ($K\text{-}mer_{RC}$) K-mer \cite{canonical}. 
\begin{equation*}
canonical~K\text{-}mer=\begin{cases}
      K\text{-}mer & \text{if $h(K\text{-}mer) < h(K\text{-}mer_{RC})$}\\
      K\text{-}mer_{RC} & \text{Otherwise}
    \end{cases}
\end{equation*}
Where $h()$ is a hash function

\textbf{Why is canonical K-mer considered in K-mer counting?}\\
During sequencing of the double-stranded DNA sequence, first, the two strands are separated and one is randomly selected to be decrypted by the machine. In other words, a DNA sequence has both original or reverse complement K-mers in different locations. Both are the same scientifically but different based on the nucleotide. Considering both as different induces errors in the frequency calculation of the K-mer counting process. Thus, the K-mer counting technique tenacious uses the canonical K-mer. 

\section{Methodology}
KmerCo is a fast approximate K-mer counting technique capable of processing billions of K-mers quickly using a small-sized Bloom Filter. It implements a CBF variant called countBF \cite{countBF}. The countBF is a two-dimensional CBF where each cell consists of many counters. The countBF performs a few arithmetic operations for fast and high performance. KmerCo takes a DNA sequence as input and produces three files as output. All the K-mers present in the DNA sequence are inserted into the countBF, hence, it can be used to determine the presence of a K-mer in the DNA sequence. The three files contain distinct, trustworthy, and erroneous K-mers. The distinct file contains the list of all distinct K-mers present in the input DNA sequence. The trustworthy file contains the list of all distinct K-mers having a frequency of more than a threshold value, say $\tau$. Whereas the erroneous file contains the list of all distinct K-mers having a frequency less than or equal to $\tau$. This section explains the working of KmerCo in detail. Table \ref{term} lists the term and notation used in the article for better understanding. 

\begin{table}    
    \begin{tabular}{|p{2cm}| p{5.7cm}|} \hline
        \centering Terms/Notations & \hspace{1.5cm} Description \\ \hline \hline 
        \hspace{0.8cm}$k_h$ & Number of hash functions \\ \hline
        \hspace{0.6cm}K-mer & $Read$ of a DNA sequence having length K\\ \hline
        \hspace{0.8cm}$K$ & Length of $read$ \\ \hline
        \vspace{0.1cm}\hspace{0.8cm}$\tau$ & Threshold value for determining the trustworthy K-mers \\ \hline 
        Distinct K-mer & Distinct K-mer present in the DNA sequence \\ \hline 
        Trustworthy K-mer & K-mer having frequency more than $\tau$ \\ \hline 
        Erroneous K-mer & K-mer having frequency less than or equal to $\tau$ \\ \hline 
    \end{tabular}
    \caption{Term/Notations used in the article and its description}
    \label{term}
\end{table}

\begin{figure}[!ht]
    \centering
    \includegraphics[width=0.45\textwidth]{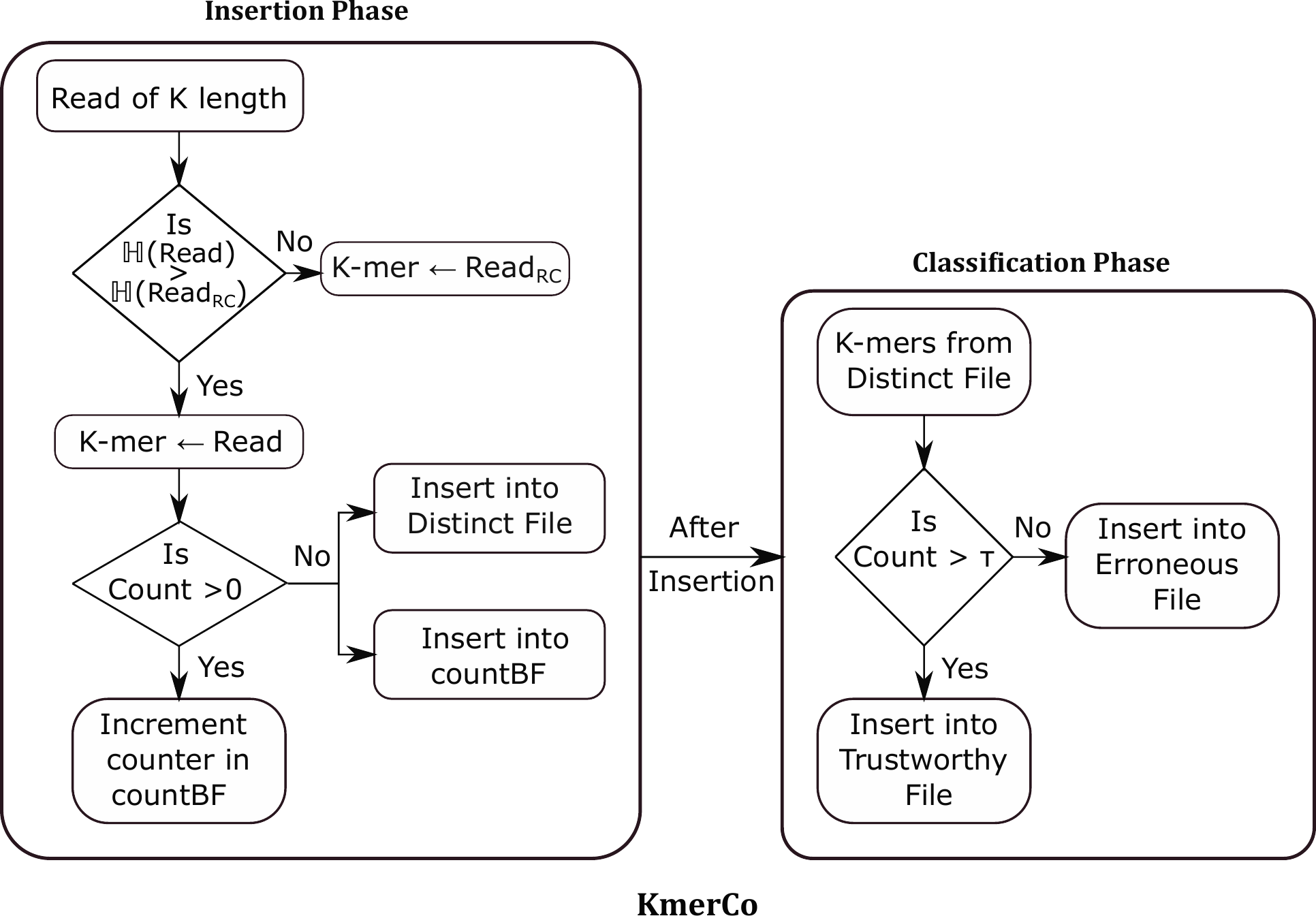}
    \caption{\textbf{Working of KmerCo}}
    \label{flowchart}
\end{figure}

Figure \ref{flowchart} illustrates the working of KmerCo. KmerCo has two phases: insertion and classification. The responsibilities of the insertion phase are extraction of $read$ from the DNA sequence file, insertion of K-mers into countBF, and distinct file. The responsibility of the classification phase is the classification of distinct K-mers into trustworthy and erroneous K-mers. In the insertion phase, first, the $Read$ of K length is extracted from the DNA sequence and obtains its reverse complement. KmerCo is a canonical K-mer counting technique which means it considers the canonical K-mers. The $Read$ and its reverse complement are hashed by a hash function, we have considered the murmur hash function \cite{murmur}. The canonical K-mer is the $Read$ or its reverse complement which has the smallest hash value. The canonical K-mer is queried to the countBF. If returns zero it means the K-mer is absent in countBF, then the K-mer is inserted into countBF. Moreover, the K-mer is encountered for the first time, hence, it is also inserted into  the distinct file. If countBF returns a non-zero value then increment the counter of the K-mer. The insertion phase completes after processing all K-mers of the DNA sequence. The output of this phase is countBF and the distinct file which are forwarded to the classification phase. In the classification phase, the K-mers are read from the distinct file and queried to the countBF. The countBF returns the frequency of the K-mer. If the frequency is more than $\tau$, then the K-mer is classified as a trustworthy K-mer and written to the trustworthy file. Otherwise, the K-mer is an erroneous K-mer and is written to the erroneous file.  The trustworthy and erroneous file can be replaced with Bloom Filter, for instance, robustBF \cite{robustbf} for fast determination of the status of the K-mer, i.e., trustworthy or erroneous.  

\begin{algorithm} 
\caption{Insertion phase of KmerCo.}
\begin{algorithmic}[1]
\Input
\\ $DNAfile$: A DNA sequence file \\
$\mathbb{C}_{x,y}$: countBF \\
$K$: Length of K-mer \\
$k_h$: Number of hash functions
\EndInput
\Output
\\ Distinct file: A file containing all the distinct K-mers present in $DNAfile$
\EndOutput
\Procedure{InsertKmerCo}{$DNAfile,~\mathbb{C}_{x,y},~K,~k_h$} 
\While{$Read~\not=~EOF$} \Comment{$EOF$: End of file}
    \State $Read$ $\leftarrow$ K-mer of length $K$
    \State $Read_{RC}$ $\leftarrow$ Reverse complement of $Read$
    \If{$\Call{QcountBF-I}{\mathbb{C}_{x,y},~K, ~Read,~Read_{RC},~k_h,~Result}=0$}
        \If{Result=0} 
                \State Insert $Read$ into Distinct file 
                \State $\Call{IcountBF}{\mathbb{C}_{x,y},~Read,~k_h}$
        \Else 
            \State Insert $Read_{RC}$ into Distinct file
            \State $\Call{IcountBF}{\mathbb{C}_{x,y},~Read_{RC},~k_h}$
        \EndIf 
    \Else
        \If{Result=0}
            \State $\Call{IcountBF}{\mathbb{C}_{x,y},~Read,~k_h}$
        \Else 
            \State $\Call{IcountBF}{\mathbb{C}_{x,y},~Read_{RC},~k_h}$
        \EndIf 
    \EndIf       
\EndWhile
\EndProcedure
\end{algorithmic}
\label{Algo1}
\end{algorithm} 
Algorithm \ref{Algo1} presents the insertion phase of KmerCo. It requires the $DNAfile$ which is the DNA sequence file, K, and $k_h$. The $Read$ of K length is extracted from the $DNAfile$. Both $Read$ and its reverse complement are passed as an argument to the $\textsc{QcountBF-I}()$ to determine whether it is present in the countBF or not. The value of the Result parameter of the $\textsc{QcountBF-I}()$ indicates the canonical K-mer. If both frequency and result are zero, then $Read$ is inserted into countBF and written to the distinct file. In case, the frequency is zero and the Result parameter is 1, then the canonical K-mer is the reverse complement of the $Read$, and it is inserted into countBF and written to the distinct file. In case, the frequency is non-zero, then increment the counter of the canonical K-mer which is determined by the Result parameter. 

\begin{algorithm} 
\caption{Classification phase of KmerCo.}
\begin{algorithmic}[1]
\Input
\\ $\mathbb{C}_{x,y}$: countBF \\
$k_h$: Number of hash functions \\
Distinct file: A file containing all the distinct K-mers present in the input DNA sequence file. 
\EndInput
\Output
\\ Trustworthy File:  A file containing K-mers having frequency more than $\tau$ \\
Erroneous File: A file containing K-mers having frequency less than or equal to $\tau$
\EndOutput
\Procedure{QueryKmerCo}{$\mathbb{C}_{x,y},~k_h$, Distinct file} 
\While{Distinct file}
    \State K-mer $\leftarrow$ K-mer read from Distinct file
        \If{$\Call{QcountBF-II}{\mathbb{C}_{x,y}, \text{K-mer}, k_h}~>~\tau$}
            \State Insert into Trustworthy File
        \Else
            \State Insert into Erroneous File 
    \EndIf       
\EndWhile
\EndProcedure
\end{algorithmic}
\label{Algo2}
\end{algorithm} 
Algorithm \ref{Algo2} illustrates the classification phase of KmerCo. Its inputs are $\mathbb{C}_{x,y}$ the countBF containing all distinct K-mers present in the DNA sequence inserted into it, $k_h$, and the distinct file containing the list of all distinct K-mers. The K-mers are read from the distinct file and queried to $\textsc{QcountBF-II}()$. The $\textsc{QcountBF-II}()$ returns the frequency of the K-mer. If the K-mer has a frequency of more than $\tau$ then it is a trustworthy K-mer; otherwise, erroneous K-mer. Based on the classification, the K-mer is written to the trustworthy or erroneous file. 

\begin{algorithm} 
\caption{Insertion operation of countBF}
\begin{algorithmic}[1]
\Input
\\ $\mathbb{C}_{x,y}$: countBF \\
K-mer: A $Read$ of length $K$ \\
$k_h$: Number of hash functions
\EndInput
\Procedure{IcountBF}{$\mathbb{C}_{x,y}$, K-mer, $k_h$}
\For{$a:~ 1~ to~ k_h$}
    \State $h$ $\leftarrow$ $\mathbb{H}_a(Read)$ \Comment{$\mathbb{H}_a()$ is a hash function}
    \State i $\leftarrow$ $h$\%x, j $\leftarrow$ $h$\%y, l $\leftarrow$ $h$\%$\eta$ \Comment{$\eta$: number of counters in \hspace*{\fill}each cell of $\mathbb{C}_{x,y}$} 
    \State value $\leftarrow$ $\mathbb{C}_{i,j}~\land~\mathcal{M}^e_l$ \Comment{$\mathcal{M}^e_l$ is the extract mask}
    \State value $\leftarrow$ $\mathcal{C}_l\gg (\alpha*l)$ \Comment{$\alpha$: counter bit length}
    \State value $\leftarrow$ value+1 
    \If{value=MAX}
        \State Counter Overflow
        \State \Return
    \EndIf
    \State value $\leftarrow$ value $ \ll~(\alpha*l)$ \Comment{Left-shift bit operation}
    \State value $\leftarrow$ $\mathbb{C}_{i,j}~\land~\mathcal{M}^r_l$ \Comment{$\mathcal{M}^r_l$ is the reset mask}
    \State $\mathbb{C}_{i,j}=\mathbb{C}_{i,j}~\lor$ value
\EndFor    
\EndProcedure
\end{algorithmic}
\label{Algo3}
\end{algorithm} 

Algorithm \ref{Algo3} demonstrates the insertion operation of countBF. The K-mer is hashed by a hash function. The modulo operation of hash value with the dimension of $\mathbb{C}_{x,y}$ and the number of counters per cell provides the required cell and counter location. Then the whole cell value is extracted by performing the $AND$ operation with the predefined extract mask, $\mathcal{M}^e_l$. The right shift operation is performed to obtain only the required counter value. The counter value is incremented and verified if the value is the MAX value permitted in the counter. If yes, then the counter overflows; hence, the operation is terminated. Otherwise, the incremented value is left shifted and performs $AND$ operation with the predefined reset mask to obtain the new counter value with respect to the cell location. Then the new cell value is inserted into the countBF using $OR$ operation with the old cell value. This whole procedure is repeated for $k_h$ times.

Noteworthy, the insertion operation causes an overflow issue in some cases. However, it does not affect the classification of K-mers: trustworthy and erroneous K-mers. On the contrary, it can affect the classification process if $\tau=2^\alpha$ but we always choose $\tau<< 2^\alpha$. 

\begin{algorithm} 
\caption{Query-I operations of countBF.}
\begin{algorithmic}[1]
\Input
\\ $\mathbb{C}_{x,y}$: countBF \\
$Read$: Query K-mer\\
$Read_{RC}$: Reverse complement of $Read$ \\
$k_h$: Number of hash functions
\EndInput
\Output
\\Frequency \\
 Result
\begin{equation*}
    Result \leftarrow \begin{cases}
      0 & \text{if $Read$ is selected} \\
      1 & \text{if $Read_{RC}$ is selected}
    \end{cases}
    \end{equation*}
\EndOutput
\Procedure{QcountBF-I}{$\mathbb{C}_{x,y},~Read,~Read_{RC},~k_h,~Result$} 
    \For{$a:~ 1~ to~ k_h$}
    \State $h$ $\leftarrow$ $\mathbb{H}_a(Read)$ \Comment{$\mathbb{H}_a()$ is a hash function}
    \State $h_1$ $\leftarrow$ $\mathbb{H}_a(Read_{RC})$
    \If{$h~<~h_1$}       
            \State Result $\leftarrow$ 0
    \Else 
        \State $h$ $\leftarrow$ $h_1$
        \State  Result $\leftarrow$ 1
    \EndIf 
        \State i $\leftarrow$ $h$\%x, j $\leftarrow$ $h$\%y, l $\leftarrow$ $h$\%$\eta$ \Comment{$\eta$: number of counters in \hspace*{\fill}each cell} 
        \State value $\leftarrow$ $\mathbb{C}_{x,y} ~\wedge~~~ \mathcal{M}^e_l$ \Comment{$\mathcal{M}^e_l$ is the extract mask}
        \State value $\leftarrow$ value $\gg~(\alpha*l)$ \Comment{$\alpha$: counter bit length}
        \If{value = 0}
            \State \Return 0
        \Else
            \State $count_a~\leftarrow$ value
        \EndIf 
    \EndFor  
    \State \Return min($count_a$) \Comment{return minimum value within count \hspace*{\fill}array}
    \EndProcedure
\end{algorithmic}
\label{Algo4}
\end{algorithm} 

\begin{algorithm} 
\caption{Query-II operations of countBF.}
\begin{algorithmic}[1]
\Input
\\ $\mathbb{C}_{x,y}$: countBF \\
K-mer: Query K-mer \\
$k_h$: Number of hash functions
\EndInput
\Output
\\ \textbf{Return:} Frequency
\EndOutput
\Procedure{QcountBF-II}{$\mathbb{C}_{x,y}$, K-mer, $k_h$}
    \For{$a:~ 1~ to~ k_h$}
        \State $h$ $\leftarrow$ $\mathbb{H}_a(K-mer)$ \Comment{$\mathbb{H}_a()$ is a hash function}
        \State i $\leftarrow$ $h$\%x, j $\leftarrow$ $h$\%y, l $\leftarrow$ $h$\%$\tau$ \Comment{$\eta$: number of counters in \hspace*{\fill} each cell} 
        \State value $\leftarrow$ $\mathbb{C}_{x,y} ~\wedge~~~ \mathcal{M}^e_l$ \Comment{$\mathcal{M}^r_l$ is the extract mask}
        \State value $\leftarrow$ value $>>~(\alpha*l)$ \Comment{$\alpha$: counter bit length}
        \If{value == 0}
            \State return 0
        \Else
            \State $count_a~\leftarrow$ value
        \EndIf
    \EndFor  
    \State \Return $\Call{min}{count_a}$ \Comment{return minimum value within count \hspace*{\fill} array}
\EndProcedure
\end{algorithmic}
\label{Algo5}
\end{algorithm} 
We propose two variants of query operations for KmerCo to optimize the execution time: $\textsc{QcountBF-I}()$ and $\textsc{QcountBF-II}()$. Algorithm \ref{Algo4} presents the $\textsc{QcountBF-I}()$ which takes the $\mathbb{C}_{x,y}$ the countBF, $Read$, $Read_{RC}$ is the reverse complement of $Read$, and $k_h$ as inputs. First, it hashes both $Read$ and $Read_{RC}$ to determine the canonical K-mer. Among the two hash values, the K-mer having the lowest hash value is the canonical K-mer which is queried to the countBF. If $Read$ is selected then the Result parameter is set to zero; otherwise one. The $\textsc{IcountBF}()$ uses the value of the Result parameter and directly inserts the $Read$ or $Read_{RC}$ without determining the canonical K-mer again. A similar procedure as in $\textsc{IcountBF}()$ is followed to obtain the counter value. If the counter value is zero, then it returns zero. Otherwise, save the counter value in an array. This procedure is followed for $k_h$ times. Then, the $\textsc{QcountBF-I}()$ returns the minimum value among the counter values. Algorithm \ref{Algo5} presents the $\textsc{QcountBF-II}()$ which takes a single K-mer as input. It returns the counter value of the K-mer. In case, the $k_h$ is greater than 1, then the $\textsc{QcountBF-II}()$ returns the minimum among the counter values. 

\section{Experiments}
We have conducted rigorous experiments to prove the supremacy of our proposed technique compared to other state-of-the-art K-mer counting techniques. KmerCo is a fast K-mer counting technique with the best performance. KmerCo takes less time for the construction of its countBF which is a classifier to classify the K-mers into trustworthy and erroneous K-mers. We have used four real datasets of different organisms of different sizes for the experiments. We have trimmed some DNA sequences from the real dataset to construct different size datasets. It helps in determining the performance of the KmerCo with big-sized datasets. We have considered two K values for experimentation: 28 and 55. This helps in showcasing the best performance of KmerCo in varying scenarios of different K length $Read$s. Other K-mer counting techniques evaluate their performance using the number of identified distinct and trustworthy K-mers which is merely a tabulation of information without any benchmark for comparison. In this paper, we have proposed a new benchmark for comparison to determine the accuracy and performance of the K-mer counting techniques. We used the Hadoop MapReduce program to determine the exact number of distinct, trustworthy, and erroneous K-mers of the datasets. We have compared KmerCo with Squeakr, BFCounter, and Jellyfish K-mer counting techniques. We have measured the performance of the techniques using data structure size, insertion time, number of insertions, inserted-to-ignored K-mer ratio, number of insertions/second, and trustworthy rate. This section provides detailed information regarding the dataset and experimentation. We have also conducted some experiments on the countBF of KmerCo to determine the counter length per cell and the number of input items for the construction of the Bloom Filter. This information is presented in the supplementary document. 
We have conducted the experiments in a low-cost Ubuntu-Desktop computer with 4GB RAM and a Core-i7 processor.   

\subsection{Dataset Description}
We have used four real datasets of different organisms, specifically mammals in our experimentation. The organisms are Loxodonta cyclotis (common name: Elephant, downloaded from \cite{elephant}), Galeopterus variegatus (common name: Sunda flying lemur, downloaded from \cite{lemur}), Microcebus murinus (common name: grey mouse lemur, downloaded from \cite{lemur1}), and Balaenoptera acutorostrata (common name: minke whale, downloaded from \cite{whale}). Table \ref{data_des} provides other details regarding the datasets. We have trimmed the real dataset to have four different size datasets. The aim is to observe the performance of KmerCo and other K-mer counting techniques in the case of different-size datasets. We have used the first word of the organism’s scientific name in the rest of the article. 

\begin{table}
    \centering
    \begin{tabular}{|p{1.7cm} | p{1cm}| p{1.6cm}| p{1.4cm}| p{1cm}|} \hline
Species & Down- load link & SRA Accession  & \#Sequences & File Size\\ \hline \hline
Loxodonta cyclotis & \cite{elephant} & SRR12606482	& 550262 & 100 MB	\\ \hline
Galeopterus variegatus & \cite{lemur} & SRR3683902 & 1622500 & 200 MB	\\ \hline
Microcebus murinus & \cite{lemur1} & SRR20563527 & 866250 & 300.1 MB	\\ \hline
Balaenoptera acutorostrata & \cite{whale} & SRR17322416	& 748250 & 400.4 MB	 \\ \hline
    \end{tabular}
    \caption{Dataset Details. \#Sequences: Number of sequences present initially in the downloaded file. File Size is in megabytes (MB) after trimming the dataset.} 
    \label{data_des}
\end{table}

\subsection{Frequency counting using Hadoop MapReduce}
The other K-mer counting techniques present only the number of distinct and trustworthy K-mers. However, these numbers do not provide any comparison of performance between the techniques. Hence, we have used the Hadoop MapReduce program for determining the exact number of distinct, trustworthy, and erroneous K-mers in the datasets. We generated K-mers of lengths 28 and 55 of the four real datasets separately in different files. These files are input into the Hadoop MapReduce program \cite{MapReduce}. After execution of the program, the output file gives the list of  distinct K-mers along with their frequency. Using the frequency, we determined the trustworthy and erroneous K-mers. This program gives no errors, thus, we can confidently use this information for comparing the performance between KmerCo and other techniques. Table \ref{hadoop28} and Table \ref{hadoop55} exhibit the total, distinct, and trustworthy 28-mers and 55-mers, respectively. The erroneous K-mers is the difference between the distinct and trustworthy K-mers, hence, it is excluded from the tables. 

\begin{table}
    \centering
    \begin{tabular}{|p{1.8cm}| p{1.5cm}| p{2cm}| p{1.7cm}|} \hline
\textbf{Dataset} & \textbf{$\#$28-mers} & \textbf{$\#$Distinct} & \textbf{$\#$Trustworthy}  \\ \hline \hline
Loxodonta & 41013058 & 32512928 & 185770 \\ \hline
Galeopterus & 74824972 & 42294113 &	581671 \\ \hline
Microcebus  & 130803722 & 43261507 & 6205318 \\ \hline
Balaenoptera & 163872472 & 38775701 & 2701406 \\ \hline
\end{tabular}
\caption{Details of 28-mers determined by the Hadoop MapReduce program. $\#$28-mers: Total number of 28-mers, $\#$Distinct: Number of distinct 28-mers, and $\#$Trustworthy: Number of trustworthy 28-mers having frequency more than $\tau=5$.}
\label{hadoop28}
\end{table}

\begin{table}
    \centering
    \begin{tabular}{|p{1.8cm}| p{1.5cm}| p{2cm}| p{1.7cm}|} \hline
\textbf{Dataset} & \textbf{$\#$55-mers} & \textbf{$\#$Distinct} & \textbf{$\#$Trustworthy} \\ \hline \hline
Loxodonta & 41013031 &	40198219 & 24629 \\ \hline
Galeopterus & 74824945 & 62985971 & 334397 \\ \hline
Microcebus  & 130803695 &	65779817 & 4633902 \\ \hline
Balaenoptera & 163872444& 79374926 & 2030213 \\ \hline

\end{tabular}
\caption{Details of 55-mers determined by the Hadoop MapReduce program. $\#$55-mers: Total number of 55-mers, $\#$Distinct: Number of distinct 55-mers, and $\#$Trustworthy: Number of trustworthy 55-mers having frequency more than $\tau=5$.}
\label{hadoop55}
\end{table}

\subsection{Dataset Analysis}
\begin{figure*}[!ht] 
\centering 
\begin{subfigure}[b!]{0.33\textwidth}
\begin{tikzpicture}
    \begin{axis}[
    width=0.99\textwidth,
    height=0.5\textwidth,
    symbolic x coords={Loxodonta,Galeopterus,Microcebus,Balaenoptera},
    ylabel={Rate},
    xlabel={Dataset},
    xtick distance=1,
    tick label style={font=\footnotesize},
    label style={font=\footnotesize},
    x tick label style={rotate=35,anchor=east},
    enlarge x limits={abs={2*(\pgfplotbarwidth+2pt)}},
    ylabel style={text width=2cm, align=center},
    enlarge x limits=0.1
    ]
\addplot [cadmiumgreen,only marks,mark=square*] table[x=dataset,y=28dis-rate]{\data};
\end{axis}
\end{tikzpicture}
\caption{Distinct 28-mer rate}
\label{fig3a}
\end{subfigure} 
\hfill
\begin{subfigure}[b!]{0.33\textwidth} 
\begin{tikzpicture}
    \begin{axis}[
    width=0.99\textwidth,
    height=0.5\textwidth,
    symbolic x coords={Loxodonta,Galeopterus,Microcebus,Balaenoptera},
    ylabel={Rate},
    xlabel={Dataset},
    xtick distance=1,
    tick label style={font=\footnotesize},
    label style={font=\footnotesize},
    x tick label style={rotate=35,anchor=east},
    enlarge x limits={abs={2*(\pgfplotbarwidth+2pt)}},
    ylabel style={text width=2cm, align=center},
    enlarge x limits=0.1
    ]
\addplot [cadmiumgreen,only marks,mark=*] table[x=dataset,y=28trust-rate]{\data};

\end{axis}
\end{tikzpicture}
\caption{Trustworthy 28-mer rate}
\label{fig3b}
\end{subfigure} 
\begin{subfigure}[b!]{0.33\textwidth} 
\begin{tikzpicture}
    \begin{axis}[
    width=0.99\textwidth,
    height=0.5\textwidth,
    symbolic x coords={Loxodonta,Galeopterus,Microcebus,Balaenoptera},
    ylabel={Rate},
    xlabel={Dataset},
    xtick distance=1,
    tick label style={font=\footnotesize},
    label style={font=\footnotesize},
    x tick label style={rotate=35,anchor=east},
    enlarge x limits={abs={2*(\pgfplotbarwidth+2pt)}},
    ylabel style={text width=2cm, align=center},
    enlarge x limits=0.1
    ]

\addplot [cadmiumgreen,only marks,mark=triangle*] table[x=dataset,y=28error-rate]{\data};

\end{axis}
\end{tikzpicture}
\caption{\scriptsize Erroneous 28-mer rate}
\label{fig3c}
\end{subfigure} 
\begin{subfigure}[b!]{0.33\textwidth} 
\begin{tikzpicture}
    \begin{axis}[
    width=0.99\textwidth,
    height=0.5\textwidth,
    symbolic x coords={Loxodonta,Galeopterus,Microcebus,Balaenoptera},
    ylabel={Rate},
    xlabel={Dataset},
    xtick distance=1,
    tick label style={font=\footnotesize},
    label style={font=\footnotesize},
    x tick label style={rotate=35,anchor=east},
    enlarge x limits={abs={2*(\pgfplotbarwidth+2pt)}},
    ylabel style={text width=2cm, align=center},
    enlarge x limits=0.1
    ]
\addplot [cadmiumgreen,only marks,mark=square*] table[x=dataset,y=55dis-rate]{\data};

\end{axis}
\end{tikzpicture}
\caption{\scriptsize Distinct 55-mer rate}
\label{fig3d}
\end{subfigure} 
\hfill
\begin{subfigure}[b!]{0.33\textwidth} 
\begin{tikzpicture}
    \begin{axis}[
    width=0.99\textwidth,
    height=0.5\textwidth,
    symbolic x coords={Loxodonta,Galeopterus,Microcebus,Balaenoptera},
    ylabel={Rate},
    xlabel={Dataset},
    xtick distance=1,
    tick label style={font=\footnotesize},
    label style={font=\footnotesize},
    x tick label style={rotate=35,anchor=east},
    enlarge x limits={abs={2*(\pgfplotbarwidth+2pt)}},
    ylabel style={text width=2cm, align=center},
    enlarge x limits=0.1
    ]
\addplot [cadmiumgreen,only marks,mark=*] table[x=dataset,y=55trust-rate]{\data};
\end{axis}
\end{tikzpicture}
\caption{\scriptsize Trustworthy 55-mer rate}
\label{fig3e}
\end{subfigure} 
\begin{subfigure}[b!]{0.33\textwidth} 
\begin{tikzpicture}
    \begin{axis}[
    width=0.99\textwidth,
    height=0.5\textwidth,
    symbolic x coords={Loxodonta,Galeopterus,Microcebus,Balaenoptera},
    ylabel={Rate},
    xlabel={Dataset},
    xtick distance=1,
    tick label style={font=\footnotesize},
    label style={font=\footnotesize},
    x tick label style={rotate=35,anchor=east},
    enlarge x limits={abs={2*(\pgfplotbarwidth+2pt)}},
    ylabel style={text width=2cm, align=center},
    enlarge x limits=0.1
    ]
\addplot [cadmiumgreen,only marks,mark=triangle*] table[x=dataset,y=55error-rate]{\data};
\end{axis}
\end{tikzpicture}
\caption{\scriptsize Erroneous 55-mer rate}
\label{fig3f}
\end{subfigure}

\caption{Analysis of various datasets based on (a) Distinct 28-mer rate, (b) Trustworthy 28-mer rate, (c) Erroneous 28-mer rate, (d) Distinct 55-mer rate, (e) Trustworthy 55-mers rate, and (f) Erroneous 55-mer rate.}
\label{fig3}
\end{figure*}
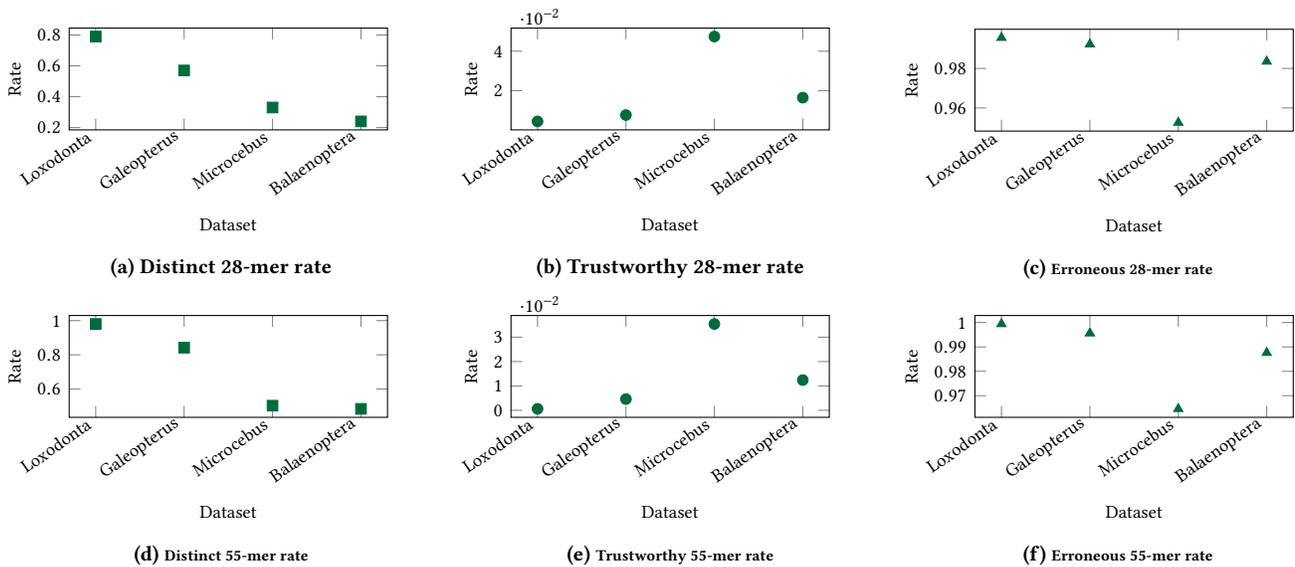

This section provides the analysis of the datasets based on the ratio of the number of distinct, trustworthy, and  erroneous K-mers to total K-mers as illustrated by Figure \ref{fig3}. 

\begin{equation*}
    Distinct ~\text{K-mer}~ rate=\frac{|Distinct~\text{K-mers}~|}{|Total~\text{K-mers}~|}
\end{equation*}
\begin{equation*}
    Trustworthy ~\text{K-mer}~ rate=\frac{|Trustworthy~\text{K-mers}~|}{|Total~\text{K-mers}|}
\end{equation*}
\begin{equation*}
  Erroneous~\text{K-mer}~ rate=\frac{|Erroneous~\text{K-mers}|}{|Total~\text{K-mers}~|}  
\end{equation*}

\pgfplotstableread[row sep=\\,col sep=&]{ 
dataset & 28dis-rate & 28trust-rate & 28error-rate & 55dis-rate & 55trust-rate & 55error-rate\\
Loxodonta & 0.79 & 0.0045 & 0.9955 & 0.9801 & 0.0006 & 0.9994 \\
Galeopterus & 0.57 & 0.0077 & 0.9922 &  0.8418 & 0.0046 &  0.9956 \\
Microcebus & 0.33 & 0.0474 & 0.9526 & 0.5029 & 0.0354 &  0.9646 \\
Balaenoptera & 0.24 & 0.0165 & 0.9835 & 0.4844 & 0.0124 & 0.9876 \\
}\data

Figure \ref{fig3a} presents the Distinct 28-mer rate where the rate decreases with an increase in dataset size. The Loxodonta dataset has the highest rate whereas the Balaenoptera dataset has the least rate which is obvious as the dataset being the lowest and highest size, respectively. Figure \ref{fig3b} highlights the Trustworthy 28-mer rate where the rate increases with an increase in dataset size with the exception of the Microcebus dataset. The Microcebus has the highest ratio whereas Loxodonta has the least ratio. Obviously, the contrary pattern will be followed in the case of an Erroneous 28-mer rate as shown in Figure \ref{fig3c}. The Loxodonta dataset has the highest rate whereas the Microcebus dataset has the least rate. Figure \ref{fig3d}, Figure \ref{fig3e}, and Figure \ref{fig3f} illustrate the Distinct, Trustworthy, and Erroneous 55-mer rate, respectively. The 55-mer dataset follows the same pattern as observed in the case of the 28-mer dataset. Overall, it is observed that the Loxodonta dataset has the highest distinct and erroneous K-mers with the least trustworthy K-mers. The Balaenoptera dataset has the lowest distinct K-mers in spite of having the highest dataset size. The Microcebus dataset has the highest trustworthy and least erroneous K-mers.

\subsection{Experimental Results}
This section provides details regarding the experimentation performed on KmerCo. The KmerCo is compared with other K-mer counting techniques: Squeakr, BFCounter, and Jellyfish. The Squeakr \cite{Squeakr} (code downloaded from \cite{Squeakrcode}) is a Bloom Filter-based technique, specifically, it implements Counting Quotient Filter (CQF) \cite{CQF}. The BFCounter \cite{BFCounter} (code downloaded from \cite{BFCountercode}) implements both standard Bloom Filter and hashtable. Jellyfish2 \cite{jellyfish} (code downloaded from \cite{jellyfishcode}) is a hashtable-based K-mer counting technique. 

\pgfplotstableread[row sep=\\,col sep=&]{ 
dataset & 28-KmerCo & 28-Squeakr & 28-BFCounter & 28-Jellyfish & 55-KmerCo & 55-Squeakr & 55-BFCounter & 55-Jellyfish \\
Loxodonta & 18 & 8.0 & 46.92 & 2368 & 18 & 0.0176 & 41.79 & 2368\\
Galeopterus & 32.6 & 32.0 & 86.75 & 4608 & 32.6 & 32 & 81.07 & 4608\\
Microcebus & 56.45 & 512 & 206.53 & 4608 & 56.45 & 256 & 190.26 & 4608\\
Balaenoptera & 70.92 & 128 & 186.17 & 8960 & 70.92 & 128 & 176.84 & 8960\\
}\mem

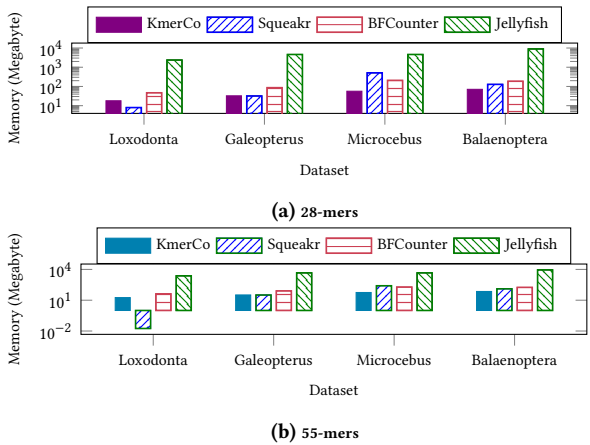
\begin{figure}[!ht] 
\centering
\begin{subfigure}{0.99\columnwidth} 
\begin{tikzpicture}
\begin{axis}[
    ybar,
    legend style={at={(0.5,1)}, anchor=south,legend columns=4,legend cell align=left,font=\scriptsize},
    xtick=data,
    enlarge x limits=0.2,
    height=0.3\textwidth,
    width=0.99\textwidth,
    bar width=2mm,
    tick label style={font=\scriptsize},
    legend style={font=\scriptsize},
    label style={font=\scriptsize},
    ylabel={Memory (Megabyte)},
    xlabel={Dataset},
    ymode=log,
    area legend,
    symbolic x coords={Loxodonta,Galeopterus,Microcebus,Balaenoptera},
    ]
\addplot+[violet] table[x=dataset,y=28-KmerCo]{\mem};
\addplot[postaction={blue,pattern=north east lines,semithick,pattern color=blue}] table[x=dataset,y=28-Squeakr]{\mem};
\addplot[myred,postaction={myred,pattern=horizontal lines,semithick,pattern color=myred}] table[x=dataset,y=28-BFCounter]{\mem};
\addplot[postaction={ao,pattern=north west lines,semithick,pattern color=ao}] table[x=dataset,y=28-Jellyfish]{\mem};
\legend{KmerCo,Squeakr,BFCounter,Jellyfish} 
\end{axis}
\end{tikzpicture}
\caption{\scriptsize 28-mers}
\label{figDS28}
\end{subfigure} 
\begin{subfigure}{0.99\columnwidth} 
\begin{tikzpicture}
\begin{axis}[
    ybar,
    legend style={at={(0.5,1)}, anchor=south,legend columns=4,legend cell align=left,font=\scriptsize},
    xtick=data,
    enlarge x limits=0.2,
    height=0.3\textwidth,
    width=0.99\textwidth,
    bar width=2mm,
    tick label style={font=\scriptsize},
    legend style={font=\scriptsize},
    label style={font=\scriptsize},
    ylabel={Memory (Megabyte)},
    xlabel={Dataset},
    ymode=log,
    area legend,
    symbolic x coords={Loxodonta,Galeopterus,Microcebus,Balaenoptera},
    ]
\addplot+[mygreen] table[x=dataset,y=55-KmerCo]{\mem};
\addplot[blue,postaction={ao,pattern=north east lines,semithick,pattern color=blue}] table[x=dataset,y=55-Squeakr]{\mem};
\addplot[myred,postaction={myred,pattern=horizontal lines,semithick,pattern color=myred}] table[x=dataset,y=55-BFCounter]{\mem};
\addplot[postaction={ao,semithick,pattern=north west lines, pattern color=ao}] table[x=dataset,y=55-Jellyfish]{\mem};
\legend{KmerCo,Squeakr,BFCounter,Jellyfish} 
\end{axis}
\end{tikzpicture}
\caption{\scriptsize 55-mers }
\label{figDS55}
\end{subfigure} 
\caption{Comparison of the memory footprint of the data structure in megabytes among KmerCo, Squeakr, BFCounter, and Jellyfish using (a) 28-mers and (b) 55-mers of various datasets. Lower is better. }
\label{figDS}
\end{figure} 

The data structure size of KmerCo is the size of countBF. Using Equation \ref{eqm}, we get $m_{countBF}=X\times Y\times \beta$ where $\beta=64$ as each cell of countBF is unsigned long int, FPP=0.001 and $n$=|total K-mers|. In the case of Squeakr, it provides the log of estimated CQF size (say $s$) as output. Squeakr uses two CQFs: global and local. Therefore, $m_{Squeakr}=2\times 2^s$. The memory size of Squeakr also depends on the total K-mers present in the dataset. It gives an option to provide the memory size, however, instead of using the provided value Squeakr calculates the memory size from the given file. Hence, it does not provide the freedom to use a large CQF to reduce the FPP. Furthermore, providing a higher memory size than the estimated value reduces the performance of Squeakr. It reduces the number of distinct, total, and trustworthy K-mers with increasing memory size which is observed during experimentation. On the contrary, if  less memory size is provided than the estimated value then it causes segmentation faults. Overall, Squeakr gives optimal performance only in the case of the estimated memory size. BFCounter uses two data structures: standard Bloom Filter and hashtable. The Bloom Filter size is the total K-mers multiplied by the number of bits per K-mer. The default value of the number of bits per K-mer is 4 but the value can be provided by the user. We have provided 8 because the countBF counter length is 8 bits. The Bloom Filter size is 8$\times$|total K-mers| bits. The size of the hashtable is the number of slots multiplied by the counter length per hashtable slot. The number of slots is provided as output. Thus, $m_{BFCounter}=(|total~\text{K-mers}|)+(8\times |slots|)$ bytes. In the case of Jellyfish, one hashtable entry size is $2K-d+r+1$ bits. The number of entries is $2^d$ where $d=\lceil(\sqrt{|total~\text{K-mers}|})\rceil$ and $r$ is calculated from reprobe (for detail refer \cite{jellyfish}). Each entry has a counter whose length is provided by the user, we have considered 8 bytes. Therefore, $m_{Jellyfish}=2^d(2K-d+r+1)$ bytes. 

Figure \ref{figDS} depicts the comparison of KmerCo with other techniques based on data structure memory size for 28-mers (Figure \ref{figDS28}) and 55-mers (Figure \ref{figDS55}) using various datasets. The memory increases with an increase in dataset size as all techniques depend on the number of K-mers for the construction of their data structure. Jellyfish have the highest memory. KmerCo has more memory compared to Squeakr only in the case of 28-mer and 55-mer Loxodonta and Galeopterus datasets. Otherwise, it has less memory compared to other techniques for other datasets. In the case of the 28-mer Loxodonta dataset, KmerCo has 10 times more memory compared to Squeakr, and 28.92 and 2350 times less memory compared to BFCounter and Jellyfish, respectively. In the case of the 28-mer Balaenoptera dataset, KmerCo has 57.08, 115.25, and 8889.08 times less memory compared to Squeakr, BFCounter, and Jellyfish,  respectively. Similarly, KmerCo is 18 times more compared to Squeakr, and 23.79 and 2350 times less memory compared to BFCounter and Jellyfish, respectively for the 55-mer Loxodonta dataset. In the case of the 55-mer Balaenoptera dataset, KmerCo has 57.08, 105.92, and 8889.08 times less memory compared to Squeakr, BFCounter, and Jellyfish, respectively. 

\pgfplotstableread[row sep=\\,col sep=&]{ 
dataset & 28-KmerCo & 28-Squeakr & 28-BFCounter & 28-Jellyfish & 55-KmerCo & 55-Squeakr & 55-BFCounter & 55-Jellyfish\\
Loxodonta & 8.44 & 14.958 & 10.4 & 8.24 & 13.10 & 8.61 & 5.43 & 5.31\\
Galeopterus & 15.97 & 18 &  22.24 & 13.29 & 24.83 & 15.07 & 15.77 & 12.92\\
Microcebus & 30.08 & 33 & 52.74 & 17.64 & 44.79 & 25.19 & 40.2 & 17.23\\
Balaenoptera & 37.55 & 26.68 & 59.38 & 21.52 & 56.3 & 15.7 & 40.04 & 19.87\\
}\time

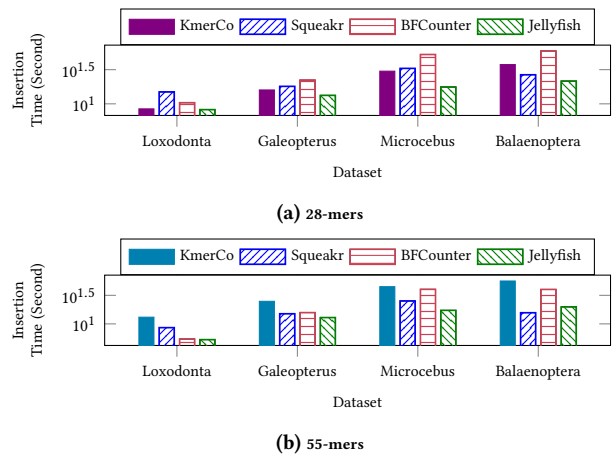
\begin{figure}[!ht] 
\centering
\begin{subfigure}{0.99\columnwidth} 
\begin{tikzpicture}
\begin{axis}[
    ybar,
    legend style={at={(0.5,1)}, anchor=south,legend columns=4,legend cell align=left,font=\footnotesize},
    xtick=data,
    enlarge x limits=0.2,
    height=0.3\textwidth,
    width=0.99\textwidth,
    bar width=2mm,
    tick label style={font=\scriptsize},
    legend style={font=\scriptsize},
    label style={font=\scriptsize},
    ylabel={Insertion Time (Second)},
    xlabel={Dataset},
    ylabel style={text width=2cm, align=center},
    ymode=log,
    area legend,
    symbolic x coords={Loxodonta,Galeopterus,Microcebus,Balaenoptera},
    ]
\addplot+[violet] table[x=dataset,y=28-KmerCo]{\time};
\addplot[postaction={blue,semithick,pattern=north east lines,pattern color=blue}] table[x=dataset,y=28-Squeakr]{\time};
\addplot[postaction={myred,semithick,pattern=horizontal lines,pattern color=myred}] table[x=dataset,y=28-BFCounter]{\time};
\addplot[postaction={ao,pattern=north west lines,semithick,pattern color=ao}] table[x=dataset,y=28-Jellyfish]{\time};
\legend{KmerCo,Squeakr,BFCounter,Jellyfish} 
\end{axis}
\end{tikzpicture}
\caption{\scriptsize 28-mers}
\label{figT28}
\end{subfigure} 
\begin{subfigure}{0.99\columnwidth} 
\begin{tikzpicture}
\begin{axis}[
    ybar,
    legend style={at={(0.5,1)}, anchor=south,legend columns=4,legend cell align=left,font=\footnotesize},
    xtick=data,
    enlarge x limits=0.2,
    height=0.3\textwidth,
    width=0.99\textwidth,
    bar width=2mm,
    tick label style={font=\scriptsize},
    legend style={font=\scriptsize},
    label style={font=\scriptsize},
    ylabel style={text width=2cm, align=center},
    ylabel={Insertion Time (Second)},
    xlabel={Dataset},
    ymode=log,
    area legend,
    symbolic x coords={Loxodonta,Galeopterus,Microcebus,Balaenoptera},
    ]
\addplot+[mygreen] table[x=dataset,y=55-KmerCo]{\time};
\addplot[postaction={blue,pattern=north east lines,semithick,pattern color=blue}] table[x=dataset,y=55-Squeakr]{\time};
\addplot[postaction={myred,pattern=horizontal lines,semithick,pattern color=myred}] table[x=dataset,y=55-BFCounter]{\time};
\addplot[postaction={ao,pattern=north west lines,semithick,pattern color=ao}] table[x=dataset,y=55-Jellyfish]{\time};
\legend{KmerCo,Squeakr,BFCounter,Jellyfish} 

\end{axis}
\end{tikzpicture}
\caption{\scriptsize 55-mers }
\label{figT55}
\end{subfigure} 
\caption{Comparison of insertion time in second among KmerCo, Squeakr, BFCounter, and Jellyfish using (a) 28-mers and (b) 55-mers of various dataset. Lower is better.}
\label{figT}
\end{figure} 

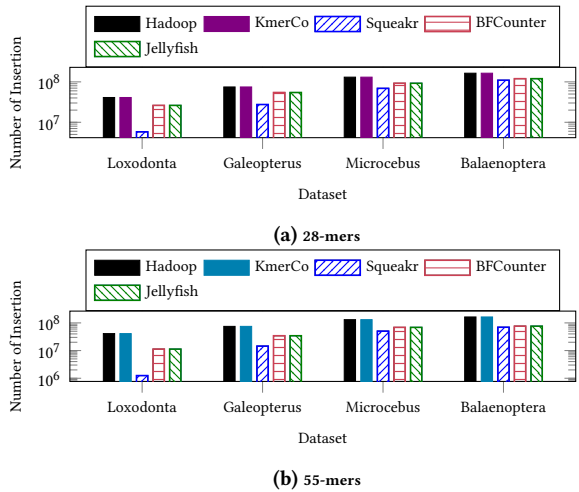
\begin{figure}[!ht] 
\centering
\begin{subfigure}{0.99\columnwidth} 
\begin{tikzpicture}
\begin{axis}[
    ybar,
    legend style={at={(0.5,1)}, anchor=south,legend columns=4,legend cell align=left,font=\footnotesize},
    xtick=data,
    enlarge x limits=0.2,
    height=0.3\textwidth,
    width=0.99\textwidth,
    bar width=1.5mm,
    tick label style={font=\scriptsize},
    legend style={font=\scriptsize},
    label style={font=\scriptsize},
    ylabel={Number of Insertion},
    xlabel={Dataset},
    area legend,
    symbolic x coords={Loxodonta,Galeopterus,Microcebus,Balaenoptera},
    ymode=log,
    ]
\addplot+[black] table[x=dataset,y=28-hadoop]{\insert};
\addplot+[violet] table[x=dataset,y=28-KmerCo]{\insert};
\addplot[postaction={blue,pattern=north east lines,semithick,pattern color=blue}] table[x=dataset,y=28-Squeakr]{\insert};
\addplot[postaction={myred,pattern=horizontal lines,semithick,pattern color=myred}] table[x=dataset,y=28-BFCounter]{\insert};
\addplot[blue,postaction={ao,pattern=north west lines,semithick,pattern color=ao}] table[x=dataset,y=28-Jellyfish]{\insert};
\legend{Hadoop,KmerCo,Squeakr,BFCounter,Jellyfish} 
\end{axis}
\end{tikzpicture}
\caption{\scriptsize 28-mers}
\label{figI28}
\end{subfigure} 
\begin{subfigure}{0.99\columnwidth} 
\begin{tikzpicture}
\begin{axis}[
    ybar,
    legend style={at={(0.5,1)}, anchor=south,legend columns=4,legend cell align=left,font=\footnotesize},
    xtick=data,
    enlarge x limits=0.2,
    height=0.3\textwidth,
    width=0.99\textwidth,
    bar width=1.5mm,
    tick label style={font=\scriptsize},
    legend style={font=\scriptsize},
    label style={font=\scriptsize},
    ylabel={Number of Insertion},
    xlabel={Dataset},
    area legend,
    symbolic x coords={Loxodonta,Galeopterus,Microcebus,Balaenoptera},
    ymode=log,
    ]
\addplot+[black] table[x=dataset,y=55-hadoop]{\insert};
\addplot+[mygreen] table[x=dataset,y=55-KmerCo]{\insert};
\addplot[postaction={blue,pattern=north east lines,semithick,pattern color=blue}] table[x=dataset,y=55-Squeakr]{\insert};
\addplot[postaction={myred,pattern=horizontal lines,semithick,pattern color=myred}] table[x=dataset,y=55-BFCounter]{\insert};
\addplot[blue,postaction={ao,pattern=north west lines,semithick,pattern color=ao}] table[x=dataset,y=55-Jellyfish]{\insert};
\legend{Hadoop,KmerCo,Squeakr,BFCounter,Jellyfish} 

\end{axis}
\end{tikzpicture}
\caption{\scriptsize 55-mers}
\label{figI55}
\end{subfigure} 
\caption{Comparison of the number of insertions among KmerCo, Squeakr, BFCounter, and Jellyfish using (a) 28-mers and (b) 55-mers of various datasets. Closer to the Hadoop bar (black bar) is better.}
\label{figI}
\end{figure}

Figure \ref{figT} elucidates the comparison of KmerCo with other techniques using insertion time for 28-mers (Figure \ref{figT28}) and 55-mers (Figure \ref{figT55}) using various datasets. The insertion time of KmerCo excludes file writing time. In the case of 28-mers, BFCounter took the highest time whereas KmerCo took the second highest time. Whereas, KmerCo took higher time than other techniques in 55-mer datasets. For the 28-mer Loxodonta dataset, KmerCo took 6.518 sec and 1.96 sec less time compared to Squeakr and BFCounter, respectively, but 0.2 sec more than Jellyfish. For the 28-mer Balaenoptera dataset, KmerCo took 10.87 sec and 16.03 sec more time compared to Squeakr and Jellyfish, respectively, but 21.83 sec  less than BFCounter. In the case of the 55-mer Loxodonta dataset, KmerCo took 4.49 sec, 7.67 sec, and 7.79 sec more than Squeakr, BFCounter, and Jellyfish, respectively; for the 55-mer Balaenoptera dataset, KmerCo took 40.6 sec, 16.26 sec, and 36.43 sec more than Squeakr, BFCounter, and Jellyfish, respectively. Overall, BFCounter and KmerCo took the highest time in the case of 28-mer and 55-mer datasets, respectively. On the contrary, Jellyfish took the lowest time in all datasets. The reason for KmerCo taking higher time is it inserts all K-mers whereas other techniques insert fewer K-mers as shown in Figure \ref{figI}. Another reason is that others are compromising and occupying large memory to have low insertion time as adduced by Figure \ref{figDS}.

\pgfplotstableread[row sep=\\,col sep=&]{ 
dataset & 28-hadoop & 28-KmerCo & 28-Squeakr & 28-BFCounter & 28-Jellyfish & 55-hadoop & 55-KmerCo & 55-Squeakr & 55-BFCounter & 55-Jellyfish\\
Loxodonta & 41013058 & 41013058 & 5804348 & 26154855 & 26154855 & 41013031 & 41013031 & 1249585 & 11297872 & 11297872\\
Galeopterus & 74824972 & 74824972 & 27523425 & 54566563 & 54566563 & 74824945 & 74824945 & 14490601 & 34380312 & 34380312\\
Microcebus & 130803722 & 130803722 & 69097995 & 92333496 & 92333496 & 130803695 & 130803695 & 50336946 & 68923817 & 68923817\\
Balaenoptera & 163872472 & 163872472 & 109746553 & 120065000 & 120065000 & 163872444 & 163872445 & 69674560 & 76257500 & 76257500\\
}\insert

Figure \ref{figI} interprets the number of insertions of 28-mers (Figure \ref{figI28}) and 55-mers (Figure \ref{figI55}) of the datasets by the techniques. The Hadoop (black bar) represents the total number of K-mers in the dataset as determined by the Hadoop MapReduce program. Hence, a bar closer to the Hadoop bar is better. KmerCo is the same as Hadoop in all datasets; whereas other techniques are less than Hadoop. Squeakr, BFCounter, and Jellyfish insert approximately 35 million, 15 million, and 15 million, respectively, fewer 28-mers in the Loxodonta dataset. As BFCounter, and Jellyfish are both hashtable-based K-mer counting techniques they insert the same number of K-mers. Similarly, Squeakr, BFCounter, and Jellyfish insert approximately 54 million, 44 million, and 44 million, respectively, fewer 28-mers in the Balaenoptera dataset. The Squeakr, BFCounter, and Jellyfish insert approximately 40 million, 30 million, and 30 million fewer 55-mers of the Loxodonta dataset. In the case of the Balaenoptera dataset, Squeakr, BFCounter, and Jellyfish insert 94 million, 88 million, and 88 million fewer 55-mers, respectively. 

\pgfplotstableread[row sep=\\,col sep=&]{ 
dataset & 28-Squeakr & 28-BFCounter & 28-Jellyfish & 55-Squeakr & 55-BFCounter & 55-Jellyfish\\
Loxodonta & 6.065919893 & 0.5680858487 & 0.5680858487 & 31.82132148 & 2.630155396 & 2.630155396\\
Galeopterus & 1.718592326 & 0.3712604915 & 0.3712604915 & 4.163688173 & 1.176389353 & 1.176389353\\
Microcebus & 0.8930176194 & 0.4166443129 & 0.4166443129 & 1.598562396 & 0.8978010896 & 0.8978010896\\
Balaenoptera & 0.4931901506 & 0.3648646317 & 0.3648646317 & 1.351969571 & 1.148935449 & 1.148935449\\
}\inig

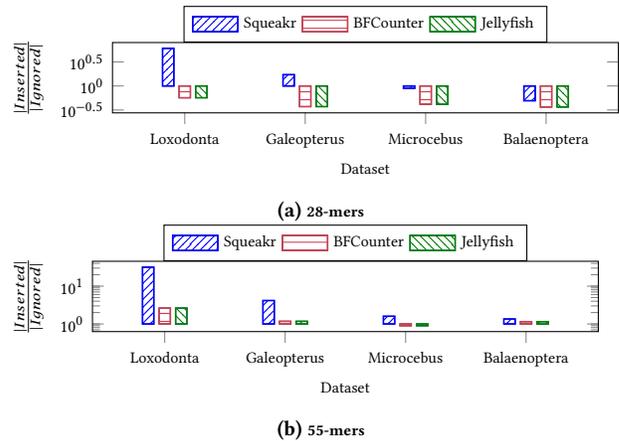
\begin{figure}[!ht] 
\centering
\begin{subfigure}{0.99\columnwidth} 
\begin{tikzpicture}
\begin{axis}[
    ybar,
    legend style={at={(0.5,1)}, anchor=south,legend columns=4,legend cell align=left,font=\footnotesize},
    xtick=data,
    enlarge x limits=0.2,
    height=0.3\textwidth,
    width=0.99\textwidth,
    bar width=1.5mm,
    tick label style={font=\scriptsize},
    legend style={font=\scriptsize},
    label style={font=\scriptsize},
    ylabel={$\frac{|Inserted|}{|Ignored|}$},
    xlabel={Dataset},
    area legend,
    symbolic x coords={Loxodonta,Galeopterus,Microcebus,Balaenoptera},
    ymode=log,
    ]
\addplot[postaction={blue,pattern=north east lines,semithick,pattern color=blue}] table[x=dataset,y=28-Squeakr]{\inig};
\addplot[postaction={myred,pattern=horizontal lines,semithick,pattern color=myred}] table[x=dataset,y=28-BFCounter]{\inig};
\addplot[blue,postaction={ao,pattern=north west lines,semithick,pattern color=ao}] table[x=dataset,y=28-Jellyfish]{\inig};
\legend{Squeakr,BFCounter,Jellyfish} 
\end{axis}
\end{tikzpicture}
\caption{\scriptsize 28-mers}
\label{figII28}
\end{subfigure} 
\begin{subfigure}{0.99\columnwidth} 
\begin{tikzpicture}
\begin{axis}[
    ybar,
    legend style={at={(0.5,1)}, anchor=south,legend columns=4,legend cell align=left,font=\footnotesize},
    xtick=data,
    enlarge x limits=0.2,
    height=0.3\textwidth,
    width=0.99\textwidth,
    bar width=1.5mm,
    tick label style={font=\scriptsize},
    legend style={font=\scriptsize},
    label style={font=\scriptsize},
    ylabel={$\frac{|Inserted|}{|Ignored|}$},
    xlabel={Dataset},
    area legend,
    symbolic x coords={Loxodonta,Galeopterus,Microcebus,Balaenoptera},
    ymode=log,
    ]
\addplot[postaction={blue,pattern=north east lines,semithick,pattern color=blue}] table[x=dataset,y=55-Squeakr]{\inig};
\addplot[postaction={myred,pattern=horizontal lines,semithick,pattern color=myred}] table[x=dataset,y=55-BFCounter]{\inig};
\addplot[blue,postaction={ao,pattern=north west lines,semithick,pattern color=ao}] table[x=dataset,y=55-Jellyfish]{\inig};
\legend{Squeakr,BFCounter,Jellyfish} 

\end{axis}
\end{tikzpicture}
\caption{\scriptsize 55-mers }
\label{figII55}
\end{subfigure} 
\caption{Comparison of inserted-to-ignored K-mer ratio among Squeakr, BFCounter, and Jellyfish using (a) 28-mers and (b) 55-mers of various datasets. KmerCo has zero ratios in all datasets. Positive and close to zero is better.}
\label{figII}
\end{figure} 

Figure \ref{figII} explicates the comparison of KmerCo with other techniques based on inserted-to-ignored K-mer ratio of 28-mers (Figure \ref{figII28}) and 55-mers (Figure \ref{figII55}) of the datasets. The ratio is $\frac{|Inserted~\text{K-mers}|}{|Ignored~\text{K-mers}|}$. More than 1 means more K-mers are ignored compared to inserted K-mers. KmerCo inserts all K-mers in all datasets, hence, the ratio is 0 in all datasets. Squeakr ignored the highest number of 28-mers and 55-mers of the Loxodonta dataset compared to others. Squeakr has ignored more 28-mers of the Loxodonta and Galeopterus datasets and ignored more 55-mers of all datasets compared to the inserted K-mers. Considering the BFCounter and Jellyfish, they have more ignored 55-mers in all datasets. 

\pgfplotstableread[row sep=\\,col sep=&]{ 
dataset & 28-KmerCo & 28-Squeakr & 28-BFCounter & 28-Jellyfish & 55-KmerCo & 55-Squeakr & 55-BFCounter & 55-Jellyfish\\
Loxodonta & 4859367.062  & 388043.0539 & 2514889.904 & 3174132.888 & 3130765.725 & 145131.8235 & 2080639.411 & 2127659.51\\
Galeopterus & 4685345.773  & 1529079.167 & 2453532.509 & 4105836.193 & 3013489.529 & 961552.8202 & 2180108.561 & 2661014.861\\
Microcebus & 4348527.992  & 2093878.636 & 1750729.92 & 5234325.17 & 2920377.205 & 1998290.83 & 1714522.811 & 4000221.532	\\
Balaenoptera & 4364113.768  & 4113439.018 & 2021977.097 & 5579228.625 & 2910700.622 & 4437870.064 & 1904532.967 & 3837820.835\\
}\ins

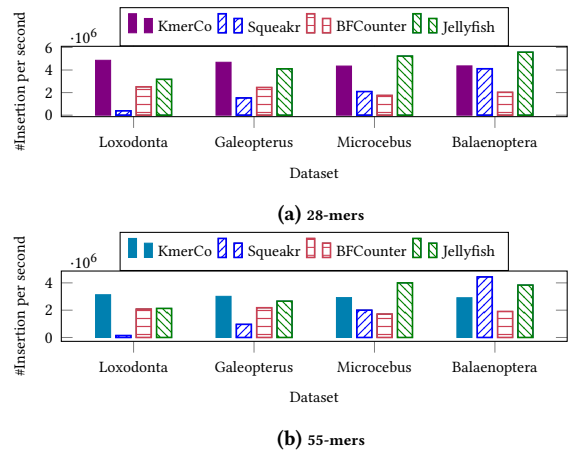
\begin{figure}[!ht] 
\centering
\begin{subfigure}{0.99\columnwidth} 
\begin{tikzpicture}
\begin{axis}[
    ybar,
    legend style={at={(0.5,1)}, anchor=south,legend columns=4,legend cell align=left,font=\footnotesize},
    xtick=data,
    enlarge x limits=0.2,
    height=0.3\textwidth,
    width=0.99\textwidth,
    bar width=2mm,
    tick label style={font=\scriptsize},
    legend style={font=\scriptsize},
    label style={font=\scriptsize},
    ylabel={\#Insertion per second},
    xlabel={Dataset},
    symbolic x coords={Loxodonta,Galeopterus,Microcebus,Balaenoptera},
    ]
\addplot+[violet] table[x=dataset,y=28-KmerCo]{\ins};
\addplot[postaction={blue,pattern=north east lines,semithick,pattern color=blue}] table[x=dataset,y=28-Squeakr]{\ins};
\addplot[postaction={myred,pattern=horizontal lines,semithick,pattern color=myred}] table[x=dataset,y=28-BFCounter]{\ins};
\addplot[blue,postaction={ao,pattern=north west lines,semithick,pattern color=ao}] table[x=dataset,y=28-Jellyfish]{\ins};
\legend{KmerCo,Squeakr,BFCounter,Jellyfish} 
\end{axis}
\end{tikzpicture}
\caption{\scriptsize 28-mers }
\label{figIS28}
\end{subfigure} 
\begin{subfigure}{0.99\columnwidth} 
\begin{tikzpicture}
\begin{axis}[
    ybar,
    legend style={at={(0.5,1)}, anchor=south,legend columns=4,legend cell align=left,font=\footnotesize},
    xtick=data,
    enlarge x limits=0.2,
    height=0.3\textwidth,
    width=0.99\textwidth,
    bar width=2mm,
    tick label style={font=\scriptsize},
    legend style={font=\scriptsize},
    label style={font=\scriptsize},
    ylabel={\#Insertion per second},
    xlabel={Dataset},
    symbolic x coords={Loxodonta,Galeopterus,Microcebus,Balaenoptera},
    ]
\addplot+[mygreen] table[x=dataset,y=55-KmerCo]{\ins};
\addplot[postaction={blue,pattern=north east lines,semithick,pattern color=blue}] table[x=dataset,y=55-Squeakr]{\ins};
\addplot[postaction={myred,pattern=horizontal lines,semithick,pattern color=myred}] table[x=dataset,y=55-BFCounter]{\ins};
\addplot[blue,postaction={ao,pattern=north west lines,semithick,pattern color=ao}] table[x=dataset,y=55-Jellyfish]{\ins};
\legend{KmerCo,Squeakr,BFCounter,Jellyfish} 
\end{axis}
\end{tikzpicture}
\caption{\scriptsize 55-mers }
\label{figIS55}
\end{subfigure} 
\caption{Comparison of the number of (a) 28-mers and (b) 55-mers inserted per second among KmerCo, Squeakr, BFCounter, and Jellyfish using various datasets. Higher is better. \#Insertion: Number of insertions.}
\label{figIS}
\end{figure} 

Figure \ref{figIS} represents the number of insertions per second for 28-mers (Figure \ref{figIS28}) and 55-mers (Figure \ref{figIS55}) of the datasets. As presented in Figure \ref{figT} although KmerCo took more insertion time, it has more insertions per second than Squeakr and BFCounter. KmerCo has the highest number of insertions per second in the case of both 28-mer and 55-mer of the Loxodonta and Galeopterus dataset; while Squeakr has the lowest. Jellyfish have the highest number of insertions per second in both 28-mer and 55-mer of Microcebus and 55-mer of Balaenoptera dataset; albeit KmerCo is the second highest in these datasets. Overall, KmerCo has a good performance with respect to the number of insertions per second. 

\pgfplotstableread[row sep=\\,col sep=&]{ 
dataset & 28-KmerCo & 28-Squeakr & 28-BFCounter & 28-Jellyfish & 55-KmerCo & 55-Squeakr & 55-BFCounter & 55-Jellyfish\\
Loxodonta & 0.0079  & -0.0006 & -0.001 & -0.001 & 0.0116 & -0.0004 & -0.0001 & -0.0001\\
Galeopterus & 0.0041 & -0.0072 & -0.0013 & -0.0013 & 0.0088 & -0.0014 & -0.0007 & -0.0007\\
Microcebus & 0.0054  & -0.0019 & -0.0048 & -0.0048 & 0.01 & -0.0099 & -0.0043 & -0.0043\\
Balaenoptera & 0.0008 & -0.0026 & -0.003 & -0.003 & 0.0029 & -0.0071 & -0.0032 & -0.0032\\
}\error

\begin{figure}[!ht] 
\centering
\begin{subfigure}{0.99\columnwidth} 
\begin{tikzpicture}
\begin{axis}[
    legend style={at={(0.5,1)}, anchor=south,legend columns=4,legend cell align=left,font=\footnotesize},
    xtick=data,
    enlarge x limits=0.2,
    height=0.35\textwidth,
    width=0.99\textwidth,
    tick label style={font=\scriptsize},
    legend style={font=\scriptsize},
    label style={font=\scriptsize},
    ylabel={Trustworthy rate},
    xlabel={Dataset},
    symbolic x coords={Loxodonta,Galeopterus,Microcebus,Balaenoptera},
    extra y tick style={
        ymajorgrids=true, 
        ytick style={/pgfplots/major tick length=0pt,}, 
        grid style={black,            
            /pgfplots/on layer=axis foreground,
        },
    },
        extra y ticks=0,
    ]
\addplot+[violet,only marks,mark=diamond] table[x=dataset,y=28-KmerCo]{\error};
\addplot+[blue,only marks,semithick,mark=square] table[x=dataset,y=28-Squeakr]{\error};
\addplot+[myred,only marks,very thick,mark=asterisk] table[x=dataset,y=28-BFCounter]{\error};
\legend{KmerCo,Squeakr,BFCounter \& Jellyfish} 
\end{axis}
\end{tikzpicture}
\caption{\scriptsize 28-mers }
\label{figTR28}
\end{subfigure} 
\begin{subfigure}{0.99\columnwidth} 
\begin{tikzpicture}
\begin{axis}[
    legend style={at={(0.5,1)}, anchor=south,legend columns=4,legend cell align=left,font=\footnotesize},
    xtick=data,
    enlarge x limits=0.2,
    height=0.35\textwidth,
    width=0.99\textwidth,
    tick label style={font=\scriptsize},
    legend style={font=\scriptsize},
    label style={font=\scriptsize},
    ylabel={Trustworthy rate},
    xlabel={Dataset},
    symbolic x coords={Loxodonta,Galeopterus,Microcebus,Balaenoptera},
    extra y tick style={
        ymajorgrids=true, 
        ytick style={/pgfplots/major tick length=0pt,}, 
        grid style={black,            
            /pgfplots/on layer=axis foreground,
        },
    },
        extra y ticks=0,
    ]
\addplot[mygreen,only marks,semithick,mark=diamond] table[x=dataset,y=55-KmerCo]{\error};
\addplot[blue,only marks,semithick,mark=square] table[x=dataset,y=55-Squeakr]{\error};
\addplot[myred,only marks,very thick,mark=asterisk] table[x=dataset,y=55-BFCounter]{\error};
\legend{KmerCo,Squeakr,BFCounter \& Jellyfish} 

\end{axis}
\end{tikzpicture}
\caption{\scriptsize 55-mers }
\label{figTR55}
\end{subfigure} 
\caption{Comparison of trustworthy rate among KmerCo, Squeakr, BFCounter, and Jellyfish using (a) 28-mers and (b) 55-mers of various dataset. Positive and close to zero is better.}
\label{figTR}

\end{figure}
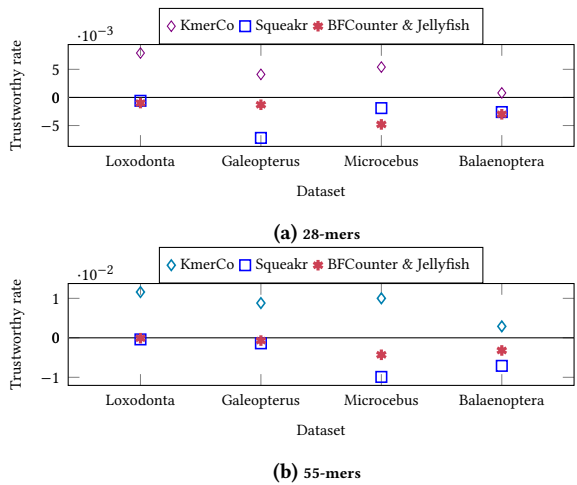 

Figure \ref{figTR} adduce the performance of KmerCo compared with other techniques based on the trustworthy rate of 28-mers (Figure \ref{figTR28}) and 55-mers (Figure \ref{figTR55}) of the datasets.
\begin{equation*}
\begin{split}
    &Trustworthy~rate=\\ &\frac{|Trustworthy~\text{K-mers}|-|Hadoop~Trustworthy~\text{K-mers}|}{|Total~\text{K-mers}|}
\end{split}
\end{equation*}
where $|Trustworthy~\text{K-mers}|$ is the trustworthy K-mers generated by the respective technique and $|Hadoop~Trustworthy~\text{K-mers}|$ is the trustworthy K-mers generated by the Hadoop MapReduce program. In the figure, close to zero means the technique correctly identifies trustworthy K-mers. More than zero, i.e., a positive trustworthy rate means the technique identifies some erroneous K-mers as trustworthy K-mer. On the contrary, less than zero, i.e., a negative trustworthy rate means some trustworthy K-mers are identified as erroneous K-mers. A positive trustworthy rate is better because the DNA assembly process ignores erroneous K-mers, hence, identifying some trustworthy K-mers as erroneous leads to loss of information. BFCounter and Jellyfish have the same trustworthy rate because they have the same number of inserted and trustworthy K-mers. In all datasets, KmerCo has a positive trustworthy rate; whereas none of the other techniques has a positive trustworthy rate in any dataset. Squeakr, BFCounter, and Jellyfish have very near to zero trustworthy rates for both 28-mer and 55-mer Loxodonta datasets. Whereas KmerCo is very close to zero trustworthy rates for both 28-mer and 55-mer Balaenoptera datasets. KmerCo has the more positive trustworthy rate in the 55-mer dataset compared to 28-mer datasets.


\section{Related Work}
The K-mer counting techniques can be broadly classified into shared and distributed memory based. The shared memory tools are further classified into hashtable-based, disk-based, and Bloom Filter-based techniques. The hashtable-based techniques use hashtable(s) to keep the counts of K-mers, for example, Jellyfish \cite{jellyfish}. The disk-based techniques have a low memory footprint but perform huge data processing using disk partitioning techniques, for example, KMC2 \cite{KMC2}, MSPKmerCounter \cite{MSPKmer}, and DSK \cite{DSK}. The Bloom Filter-based technique uses Bloom Filter as the data structure for filtering or counting K-mers, for example, KCOSS \cite{KCOSS}, Squeakr \cite{Squeakr}, and SWAPCounter \cite{SWAPCounter}. The distributed memory tools use many systems for distributed computing, for example, Kmerind \cite{Kmerind}, and Bloomfish \cite{Bloomfish}. Our proposed KmerCo is a Bloom Filter-based K-mer counting technique. Hence, this section provides a review of only Bloom Filter-based K-mer counting techniques. 

Jellyfish \cite{jellyfish} is a lightweight, multi-threaded lock-free hashtable-based K-mer counting technique. The hashtable keeps the count of the K-mers. The lock-free scheme enables parallel processing of K-mers. Each entry of the hashtable has two values: K-mer and its frequency. When the hashtable becomes saturated the data is written to disk in the form of K-mer and frequency record instead of increasing the hashtable size. 

BFCounter \cite{BFCounter} is both a Bloom Filter and hashtable-based K-mer counting technique. The hashtable keeps the count of the K-mers. It implements a standard Bloom Filter. The DNA sequence is traversed twice. In the first traversal, the K-mers are queried to the Bloom Filter, if absent it is inserted into Bloom Filter; otherwise into the hashtable. If K-mer is absent in the hashtable, insert it; otherwise, increment the counter. The second traversal is used to determine the exact frequency of the K-mers. Finally, delete all unique K-mers. The second traversal takes half the time of the first traversal as the hashtable lookup operation is faster than the insertion operation. Overall, BFCounter is slow for large datasets and the twice traversal increases the processing time. 

Mcvicar \textit{et. al.} \cite{FPGA} proposed a field programmable gate array (FPGA) and Bloom Filter-based K-mer counting technique. Bloom Filter generates small operations ideal for execution by FPGA. The CBF keeps the count of the K-mers. The technique uses 4 FPGAs and each has a Bloom Filter. The Bloom Filter implemented is CBF. The $Read$s are parsed to generate K-mers which are saved in small-size blocks and saved in a queue. From the queue, the K-mers of the blocks are hashed by the Shift-And-Xor (SAX) \cite{Ramakrishna} hash function. The hash value is forwarded to a selector which selects the FPGA that processes the K-mers. Each FPGA also has a queue which stores the K-mers which are forwarded for processing by CBF. All CBF work in parallel. The performance of the technique is independent of K. In case one hybrid memory cube (HMC) receives many operations the performance reduces due to the lack of parallel processing. This technique is best applicable in a small DNA sequence. 

Squeakr \cite{Squeakr} is an in-memory Bloom Filter-based K-mer counting technique which implements the CQF \cite{CQF}. The CQF keep the frequency of K-mers. It is a thread-based technique having two types of CQF: global and local. There is a single global CQF and each thread has a local CQF. The threads try to acquire a lock on the global CQF. The thread having the lock inserts the K-mer directly into the global CQF and others insert into the local CQF. When a local CQF becomes saturated, it is written to the global CQF. It performs a lock-free queue and implements thread-safe CQF to parallelise file parsing which enhances its ability to scale more threads. The CQF does not scale efficiently in the case of large highly skewed datasets because such datasets contain high-frequency hot-spots, i.e., regions in the sequence having many repetitive K-mers which causes excess lock contention among threads. 

SWAPCounter \cite{SWAPCounter} is a distributed Bloom Filter-based K-mer counting technique. A hashtable keeps the count of the K-mers. It implements CBF having each slot counter length $log(\theta)$ where $\theta$ is the maximum frequency among the K-mers. CBF performs the counting of K-mers. It has four components: parallel sequence I/O, K-mer extraction and distribution, K-mer filtering, and counting and statistics. The first three components are the most time-intensive tasks and the time is reduced by implementing pipelines. The first component, i.e., parallel sequence I/O, partitions the DNA sequence. In the K-mer extraction step, the DNA sequence is parsed to generate K-mers which are packed in a block. In K-mer distribution, first, the K-mer is hashed twice to determine a process and a memory location, respectively. The process is responsible for the processing of the K-mer and the K-mer is stored in the memory location. The process then queries the K-mer into the CBF.  The trustworthy K-mers are stored in a K-mer container. After filtering, a hashtable is constructed and inserts trustworthy K-mers. The message-passing interface I/O module does caching and data pooling to achieve maximum I/O performance. The technique utilises non-blocking all-to-all communication for overlapping computation and communication to enhance performance and efficacy. The computation of K-mer extraction and distribution are performed in parallel. SWAPCounter reduces the computation time by performing data compression and instruction-level optimisation to the K-mer extraction phase. Maintaining many CBFs decreases memory footprint efficiency. 

KCOSS \cite{KCOSS} is a Bloom Filter-based K-mer counting technique which implements a segmented standard Bloom. The counting of K-mer is performed by a hashtable. The segmented standard Bloom Filter is constructed by partitioning a single array into multiple Bloom Filters. The number of hash functions in KCOSS is $k_h$+1 because the first hash function determines the corresponding Bloom Filter where the K-mer is inserted or queried. KCOSS uses a shared hashtable or Bloom Filter based on K. In case $0<K\leq14$, then KCOSS uses a shared hashtable; otherwise, i.e., $K>14$ uses Bloom Filter. Along with Bloom Filter KCOSS implements two hashtables: fixed hashtable and elastic hashtable. The fixed hashtable is large in size whereas the elastic hashtable is small. The elastic hashtable is a cuckoo hashtable. The DNA sequence is partitioned into blocks and inserted into a lock-free queue. From the queue, K-mers are extracted and converted into binary format. In case $k>14$, the K-mer is queried to Bloom Filter. If the K-mer is the first occurrence, then it is stored in an overlapping sequence set. If a non-first occurrence K-mer, then it is stored in a hashtable and elastic cuckoo hashtable. When Bloom Filter returns true, then K-mer is checked in the fixed hashtable, if present then increment counter. Otherwise, check the K-mer in the cuckoo hashtable. Finally, all distinct K-mers, i.e., overlapping sequence set and the hashtables write the K-mer with its frequency to a file. The FPP of a segmented Bloom Filter is the same as a standard Bloom Filter. KCOSS  maintains many data structures; which increases the overall memory footprint. The size of the overlapping sequence set depends on the DNA sequence size, number of distinct K-mers, and number of unique K-mers. KCOSS implements shared hashtable or Bloom Filter based on K value because it believes a lower K value has fewer K-mers; however, the contrary is true, with the increase in K value the number of K-mers decreases.

\section{Conclusion}
In this paper, we proposed, KmerCo, a new efficacious and potent K-mer counting technique. It implements a low memory footprint counting Bloom Filter called countBF for low FPP and high efficiency. KmerCo has two phases: insertion and classification. In the insertion phase, countBF is constructed, i.e., K-mers are inserted into countBF and recognize distinct K-mers. In the classification phase, the distinct K-mers are queried to countBF to classify between trustworthy and erroneous K-mers. The classification is based on the user-provided threshold value. The output of KmerCo is countBF with inserted K-mers, and three files: distinct, trustworthy, and erroneous. 

We conducted a myriad of experiments to prove the dominance of KmerCo with other state-of-the-art techniques in terms of performance. The experiments are conducted using DNA sequence datasets of four different organisms, specifically mammals. The dataset was cropped to construct four different size datasets to showcase the performance with an increase in dataset size. The KmerCo was compared with Squeakr, BFCounter, and Jellyfish. KmerCo took the least memory footprint because usage of a single counting Bloom Filter, i.e., countBF, was sufficient for better and faster operation whereas other techniques implemented multiple data structures except for Jellyfish. Moreover, Jellyfish requires a large-sized hashtable for good performance. The state-of-the-art techniques sacrifice memory to lessen the insertion time. On the contrary, KmerCo maintains both the lowest memory and less insertion time. KmerCo has 57.08, 105.92, and 8889.08 times less memory compared to Squeakr, BFCounter, and Jellyfish, respectively in the 55-mers Balaenoptera dataset; whereas KmerCo took 40.6 sec, 16.26 sec, and 36.43 sec more than Squeakr, BFCounter, and Jellyfish, respectively. Another contributing factor for KmerCo’s high insertion time is the insertion of all K-mers while others have inserted fewer K-mers. In the 55-mers Balaenoptera dataset, Squeakr, BFCounter, and Jellyfish inserted approximately 94 million, 88 million, and 88 million fewer 55-mers. This leads to another comparison based on the number of K-mers inserted to the number of K-mers ignored ratio. KmerCo has a zero ratio whereas others have ignored more 55-mers in all datasets. KmerCo is the second largest in the number of insertions per second after Jellyfish as Jellyfish has the lowest insertion time but requires the largest memory compared to others. Another comparison parameter is the trustworthy rate which indicates the deviation of trustworthy K-mers identified by the state-of-the-art techniques compared to trustworthy K-mers recognized by Hadoop. KmerCo has a positive trustworthy rate whereas others have a negative rate in all datasets. A positive trustworthy rate represents that KmerCo is classifying some erroneous K-mer as trustworthy. On the other hand, a negative trustworthy rate displays that the K-mer counting technique fails to recognise some K-mers as trustworthy. It leads to an issue as trustworthy K-mers are only considered in further processing by the DNA assembly. Overall, the rigorous experiments and experimental analysis prove the dominance of KmerCo over other state-of-the-art techniques. 




\balance
\bibliographystyle{ACM-Reference-Format}
\bibliography{mybib}

\end{document}


\maketitle

\section{RESULTS}
We have conducted some additional experiments on KmerCo to evaluate its performance based on the different countBF counter lengths in bits and data structure size. These experiments are performed on the largest dataset, i.e., the Balaenoptera dataset. 

\subsection{Varying countBF counter length}
\begin{table}[H]
    \centering
    \begin{tabular}{|p{2.8cm}| p{1.8cm}| p{2.2cm}|} \hline
\textbf{Counter length (bits)} & \textbf{\#Counter} & \textbf{Wasted bits per cell} \\ \hline \hline
\hspace{1.3cm} 5 & \hspace{0.7cm} 12 & \hspace{0.9cm} 4 \\ \hline
\hspace{1.3cm} 6 & \hspace{0.7cm} 10 & \hspace{0.9cm} 4 \\ \hline
\hspace{1.3cm} 7 & \hspace{0.7cm} 9 & \hspace{0.9cm} 1 \\ \hline
\hspace{1.3cm} 8 & \hspace{0.7cm} 8 & \hspace{0.9cm} 0\\ \hline
\hspace{1.3cm} 9 & \hspace{0.7cm} 7 & \hspace{0.9cm} 1 \\ \hline
\hspace{1.3cm} 10 & \hspace{0.7cm} 6 & \hspace{0.9cm} 4\\ \hline
\hspace{1.3cm} 12 & \hspace{0.7cm} 5 & \hspace{0.9cm} 4\\ \hline
\hspace{1.3cm} 14 & \hspace{0.7cm} 4 & \hspace{0.9cm} 8\\ \hline
\hspace{1.3cm} 16 & \hspace{0.7cm} 4 & \hspace{0.9cm} 0\\ \hline
\end{tabular}
\caption{\textbf{Details of the number of counters and the wasted bits in each cell of countBF in KmerCo. \#Counter: Number of counters in each cell.}}
\label{counter}
\end{table}

In experimentation, we have considered the countBF counter length from 5 up to 16. The counter length of less than 5 is not considered because a 4-bit counter can store a maximum frequency of 15 which is a low value. Hence, the experiment is conducted by considering the counter length of more than 4 bits. Similarly, a counter length of more than 16 bits has less than 4 counters in each cell and more wasted bits. Table \ref{counter} highlights the number of counters and bits wasted per cell by varying the counter length of countBF in KmerCo. The ideal counter length is 8 and 16 with zero wasted bits. On the contrary, the counter length 14 has the highest wasted bits (with reference to Table \ref{counter}). The KmerCo have the same size but different counter lengths within a cell. All KmerCo inserted the same number of K-mers, and the inserted-to-ignored K-mer ratio is zero. Hence, in this section the comparison is presented based on insertion time, number of insertions per second, and trustworthy rate. 

\pgfplotstableread[row sep=\\,col sep=&]{
counter-len & 28time & 55time &  28error & 55error & 28-insert/sec & 55-insert/sec\\
5 & 37.65  & 56.98  & 0.0005  &  0.0015 & 4352522.497  &  2875964.286\\
6 & 37.99  & 56.81  & 0.0007  &  0.002 & 4313568.623  &  2884570.41\\
7 & 37.89  & 56.71  & 0.0007  & 0.0024  & 4324953.075  & 2889656.939 \\
8 & 37.55  & 56.3  & 0.0008  &  0.0029 & 4364113.768  &  2910700.622\\
9 & 37.57  & 56.25  & 0.001  &  0.0036 & 4361790.578  &  2913287.911\\
10 & 37.53  & 56.33  & 0.0012  &  0.0047 & 4366439.435  &  2909150.453\\
12 & 37.58  & 56.17  & 0.0015  &  0.0066 & 4360629.91  &  2917437.155\\
14 & 37.67  & 56.22  & 0.002  &  0.01 & 4350211.627  &  2914842.494\\
16 & 37.67  & 56.18  & 0.002  &  0.01 & 4350211.627  &  2916917.853\\
}\counterlen

\begin{figure}[!ht] 
\centering
\begin{subfigure}{0.99\columnwidth} 
\begin{tikzpicture}
    \begin{axis}[
    width=0.99\textwidth,
    height=0.45\textwidth,
    bar width=.5cm,
    symbolic x coords={5,6,7,8,9,10,12,14,16},
    legend style={at={(0.5,1)}, anchor=south,legend columns=2,legend cell align=left},
    ylabel={Insertion Time (second)},
    xlabel={ countBF counter length (bits)},
    xtick distance=1,
    tick label style={font=\footnotesize},
    ylabel style={text width=3cm, align=center},
    ]

\addplot [violet,thick,mark=*] table[x=counter-len,y=28time]{\counterlen};

\end{axis}
\end{tikzpicture}
\caption{28-mers}
\label{figCI28}
\end{subfigure} 
\begin{subfigure}{0.99\columnwidth} 
\begin{tikzpicture}
    \begin{axis}[
    width=0.99\textwidth,
    height=0.45\textwidth,
    bar width=.5cm,
    symbolic x coords={5,6,7,8,9,10,12,14,16},
    legend style={at={(0.5,1)}, anchor=south,legend columns=2,legend cell align=left},
    ylabel={Insertion Time (second)},
    xlabel={countBF counter length (bits)},
    xtick distance=1,
    tick label style={font=\footnotesize},
    ylabel style={text width=3cm, align=center},
    ]
    
\addplot [mygreen,thick,mark=square*] table[x=counter-len,y=55time]{\counterlen};

\end{axis}
\end{tikzpicture}
\caption{55-mers}
\label{figCI55}
\end{subfigure} time wrt counter length
\caption{\textbf{Comparison of insertion time in second among various KmerCo having different counter lengths of countBF using (a) 28-mers and (b) 55-mers Balaenoptera dataset. Lower is better.}}
\label{figCI}
\end{figure} 

Figure \ref{figCI} illustrates the comparison of KmerCo having varying counter lengths based on the insertion time in seconds using the 28-mer (Figure \ref{figCI28}) and 55-mer (Figure \ref{figCI55}) Balaenoptera dataset. KmerCo having countBF with a counter length of less than 8 bits has more insertion time compared to countBF with a counter length of more than 8 bits. In most of the cases, the insertion time increases with an increase in the counter length of countBF with exception of counter lengths 7, 8 and 10. In these cases, the insertion time decreases from the previous counter length countBF. KmerCo having countBF with a 10-bit counter has the least time followed by countBF with an 8-bit counter. But a countBF with a 10-bit counter has 4 wasted bits per cell; hence, the countBF with an 8-bit counter is better.  In the case of the 55-mer Balaenoptera dataset, the insertion time decreases with an increase in the counter length of countBF. The least time is taken by countBF with a 12-bit counter while the highest time is taken by countBF with a 5-bit counter. 

\begin{figure}[!ht] 
\centering
\begin{subfigure}{0.99\columnwidth} 
\begin{tikzpicture}
\begin{axis}[
    ybar,
    legend style={at={(0.5,1)}, anchor=south,legend columns=4,legend cell align=left,font=\footnotesize},
    xtick=data,
    enlarge x limits=0.2,
    height=0.45\textwidth,
    width=0.97\textwidth,
    bar width=3mm,
    tick label style={font=\footnotesize},
    legend style={font=\footnotesize},
    ylabel={\#Insertion per second},
    xlabel={countBF counter length (bits)},
    ylabel style={text width=2.5cm, align=center},
    area legend,
    symbolic x coords={5,6,7,8,9,10,12,14,16},
    ]
\addplot+[violet] table[x=counter-len,y=28-insert/sec]{\counterlen};

\end{axis}
\end{tikzpicture}
\caption{28-mers }
\label{figCIS28}
\end{subfigure} 
\begin{subfigure}{0.99\columnwidth} 
\begin{tikzpicture}
\begin{axis}[
    ybar,
    legend style={at={(0.5,1)}, anchor=south,legend columns=4,legend cell align=left,font=\footnotesize},
    xtick=data,
    enlarge x limits=0.2,
    height=0.45\textwidth,
    width=0.97\textwidth,
    bar width=3mm,
    tick label style={font=\footnotesize},
    legend style={font=\footnotesize},
    ylabel={$\#$Insertion per second},
    xlabel={countBF counter length (bits)},
    ylabel style={text width=2.5cm, align=center},
    area legend,
    symbolic x coords={5,6,7,8,9,10,12,14,16},
    ]
\addplot+[mygreen] table[x=counter-len,y=55-insert/sec]{\counterlen};

\end{axis}
\end{tikzpicture}
\caption{55-mers}
\label{figCIS55}
\end{subfigure} 
\caption{\textbf{Comparison of the number of (a) 28-mers and (b) 55-mers inserted per second among various KmerCo having different counter lengths of countBF using Balaenoptera dataset. Higher is better. \#Insertion: Number of insertions.}}
\label{figCIS}
\end{figure} 

Figure \ref{figCIS} illuminates the comparison among the KmerCo having different counter lengths based on the number of insertions per second using the 28-mer (Figure \ref{figCIS28}) and 55-mer (Figure \ref{figCIS55}) Balaenoptera dataset. The countBF with counter lengths less than 8 bits gives an unusual pattern as they have more counters but less number of insertions per second compared to countBF with counter lengths more than 8 bits. The countBF with a 10-bit counter exhibits the highest number of insertions per second followed by countBF with an 8-bit counter. Therefore, countBF with an 8-bit counter is better because a countBF with a 10-bit counter has 4 wasted bits per cell. In the case of 55-mer Balaenoptera dataset, the number of insertions per second increases with an increase in the counter length of countBF. The countBF with a 12-bit counter has the highest number of insertions per second. The number of insertions per second of countBF with an 8-bit counter is close to countBF with a 12-bit counter. Moreover, both countBF with 10 bits and 12 bits counters have 4 wasted bits per cell; hence countBF with an 8-bit counter is better.

\begin{figure}[!ht] 
\centering
\begin{subfigure}{0.99\columnwidth} 
\begin{tikzpicture}
\begin{axis}[
    ybar,
    legend style={at={(0.5,1)}, anchor=south,legend columns=4,legend cell align=left,font=\footnotesize},
    xtick=data,
    enlarge x limits=0.2,
    height=0.45\textwidth,
    width=0.97\textwidth,
    bar width=3mm,
    tick label style={font=\footnotesize},
    legend style={font=\footnotesize},
    ylabel={Trustworthy Rate},
    xlabel={countBF counter length (bits)},
    symbolic x coords={5,6,7,8,9,10,12,14,16},
    ]
\addplot+[violet] table[x=counter-len,y=28error]{\counterlen};

\end{axis}
\end{tikzpicture}
\caption{28-mer} 
\label{figCT28}
\end{subfigure} 
\begin{subfigure}{0.99\columnwidth} 
\begin{tikzpicture}
\begin{axis}[
    ybar,
    legend style={at={(0.5,1)}, anchor=south,legend columns=4,legend cell align=left,font=\footnotesize},
    xtick=data,
    enlarge x limits=0.2,
    height=0.45\textwidth,
    width=0.97\textwidth,
    bar width=3mm,
    tick label style={font=\footnotesize},
    legend style={font=\footnotesize},
    ylabel={Trustworthy Rate},
    xlabel={countBF counter length (bits)},
    symbolic x coords={5,6,7,8,9,10,12,14,16},
    ]
\addplot+[mygreen] table[x=counter-len,y=55error]{\counterlen};

\end{axis}
\end{tikzpicture}
\caption{55-mers}
\label{figCT55}
\end{subfigure}  
\caption{\textbf{Comparison of the trustworthy rate among various KmerCo having different counter lengths of countBF using 55-mer Balaenoptera dataset. Positive and close to zero is better.}} 
\label{figCT}
\end{figure} 

Figure \ref{figCT} highlights the comparison among various KmerCo having different counter lengths based on trustworthy rate using 28-mer (Figure \ref{figCT28}) and 55-mer (Figure \ref{figCT55}) Balaenoptera dataset. The trustworthy rate increases with the increase in the counter length in both cases, i.e., 28-mer and 55-mer. All KmerCo has a positive trustworthy rate. The trustworthy rate is better if it is closer to zero. Hence, the countBF with 5-bit has the better trustworthy rate whereas the countBF with 16-bit has the least trustworthy rate. 

Overall, KmerCo having countBF with counter length 8, 10 and 12 bits showcase better performance based on insertion time, and number of insertions per second. However, both countBF with counter lengths 10 and 12 bits waste 4 bits per cell which is a huge waste considering the whole data structure. Considering the trustworthy rate, 8-bit countBF has better performance than 10-bit or 12-bit countBF. Thus, we have considered countBF with an 8-bit counter for evaluating the performance of KmerCo using real DNA sequence datasets.

\subsection{Varying countBF data structure size}
The KmerCo’s Bloom Filter, i.e., countBF data structure size depends on the number of total K-mers present in the input datasets. We have conducted a few experiments to observe the performance of KmerCo by increasing and reducing the total number of K-mers. The counter length of countBF is 8 bits. Table \ref{tab_mem} provides details regarding the number of K-mers considered for the construction of countBF and countBF size in megabytes. All KmerCo having different countBF size inserts the same number of K-mers and the inserted-to-ignored K-mer ratio is zero. Therefore, this section presents the comparison among various KmerCo having different countBF sizes based on the insertion time, number of insertions per second, and trustworthy rate.    

\begin{table}
    \centering
\begin{tabular}{|p{2.3cm}| p{1.4cm}| p{1.4cm}| p{1.5cm}|} \hline
 & \textbf{28-mers} & \textbf{55-mers} & \textbf{CountBF size} \\ \hline \hline
Total K-mers / 8 & 20484059 & 20484055 & 9.1 \\ \hline
Total K-mers / 4 & 40968118 & 40968111 & 18 \\ \hline
Total K-mers / 2 & 81936236 & 81936222 & 36.19 \\ \hline
Total K-mers  & 163872472 & 163872445 & 70.92 \\ \hline
Total K-mers * 2 & 327744944 & 327744890 & 142.2 \\ \hline
\end{tabular}
\caption{\textbf{Details of the number of K-mers and countBF size in Megabytes by changing the number of total K-mers for construction of countBF of KmerCo using Balaenoptera dataset.}}
\label{tab_mem}
\end{table}

\pgfplotstableread[row sep=\\,col sep=&]{ 
memory & 28time & 55time & 28error & 55error & 28-insert/sec & 55-insert/sec\\
9.1 & 29.29  & 47.54  & 0.001  &  0.0388 & 5594826.63 & 3447043.437\\
18 & 33.46  & 51.83  & 0.005  & 0.031 & 4897563.419 & 3161729.597\\
36.19 & 36.14  & 54.67  & 0.002  &  0.01 & 4534379.413 & 2997483.903\\
142.2 & 38.6  & 57.46  & 0.0004  & 0.001 & 4245400.829 & 2851939.523\\
70.92 & 37.55  & 56.3  & 0.002  &  0.0029 & 4364113.768 & 2910700.622\\
}\memory

\begin{figure}[!ht] 
\centering
\begin{tikzpicture}
    \begin{axis}[
    width=0.45\textwidth,
    height=0.3\textwidth,
    symbolic x coords={2258,142,71,36,18},
    legend style={at={(0.5,1)}, anchor=south,legend columns=2,legend cell align=left},
    ylabel={Insertion Time (second)},
    xlabel={countBF size (Megabyte)},
    ylabel style={text width=2.5cm, align=center},
    xtick distance=1,
    tick label style={font=\footnotesize},
    symbolic x coords={9.1,18,36.19,70.92,142.2},
    ]
\addplot [mygreen,thick,mark=square*] table[x=memory,y=28time]{\memory};
\addplot [violet,thick,mark=*] table[x=memory,y=55time]{\memory};
\legend{28-mer,55-mer} 
\end{axis}
\end{tikzpicture}
\caption{\textbf{Comparison of insertion time in second among various KmerCo having different countBF size using 28-mer and 55-mer Balaenoptera dataset. Lower is better.}}
\label{figMI}
\end{figure} 

Figure \ref{figMI} elucidates the comparison of various KmerCo with different countBF sizes based on insertion time in seconds using 28-mer and 55-mer Balaenoptera datasets. The figure clearly highlights that the insertion time decreases with a decrease in countBF size. The 142.2 MB countBF took 38.6 sec and 57.46 sec for 28-mer and 55-mer Balaenoptera datasets, respectively. Similarly, 9.1 MB countBF took  29.29 sec and 47.54 sec for the 28-mer and 55-mer Balaenoptera datasets, respectively. On average, the insertion time increases by 2.33 sec and 2.48 sec in the 28-mer and 55-mer Balaenoptera datasets, respectively.

\begin{figure}[!ht] 
\centering
\begin{subfigure}{0.99\columnwidth} 
\begin{tikzpicture}
\begin{axis}[
    ybar,
    legend style={at={(0.5,1)}, anchor=south,legend columns=4,legend cell align=left,font=\footnotesize},
    xtick=data,
    enlarge x limits=0.2,
    height=0.45\textwidth,
    width=0.97\textwidth,
    bar width=3mm,
    tick label style={font=\footnotesize},
    legend style={font=\footnotesize},
    ylabel={$\#$Insertion per second},
    xlabel={countBF size (Megabyte)},
    ylabel style={text width=2.5cm, align=center},
    area legend,
    symbolic x coords={9.1,18,36.19,70.92,142.2},
    ]
\addplot+[violet] table[x=memory,y=28-insert/sec]{\memory};

\end{axis}
\end{tikzpicture}
\caption{28-mers }
\label{figMIS28}
\end{subfigure}  
\begin{subfigure}{0.99\columnwidth} 
\begin{tikzpicture}
\begin{axis}[
    ybar,
    legend style={at={(0.5,1)}, anchor=south,legend columns=4,legend cell align=left,font=\footnotesize},
    xtick=data,
    enlarge x limits=0.2,
    height=0.45\textwidth,
    width=0.97\textwidth,
    bar width=3mm,
    tick label style={font=\footnotesize},
    legend style={font=\footnotesize},
    ylabel={$\#$Insertion per second},
    xlabel={countBF size (Megabyte)},
    ylabel style={text width=2.5cm, align=center},
    area legend,
    symbolic x coords={9.1,18,36.19,70.92,142.2},
    ]
\addplot+[mygreen] table[x=memory,y=55-insert/sec]{\memory};

\end{axis}
\end{tikzpicture}
\caption{55-mers }
\label{figMIS55}
\end{subfigure} 
\caption{\textbf{Comparison of the number of (a) 28-mers and (b) 55-mers inserted per second among various KmerCo having different countBF size using Balaenoptera dataset. Higher is better. \#Insertion: Number of insertions.}}
\label{figMIS}
\end{figure} 

Figure \ref{figMIS} illuminate the performance of various KmerCo having different countBF size based on the number of insertions per second using 28-mer (Figure \ref{figMIS28}) and 55-mer (Figure \ref{figMIS55}) Balaenoptera dataset. The number of insertions per second increases with a decrease in KmerCo’s countBF size. All KmerCo inserted the same K-mers but insertion time decreases with decreases in KmerCo’s countBF size. On average, the number of insertions per second increases by 337356.4503 and 148775.9785 with the decrease in the KmerCo’s countBF size in the 28-mer and 55-mer Balaenoptera datasets, respectively. 

\begin{figure}[!ht] 
\centering
\begin{tikzpicture}
\begin{axis}[
    legend style={at={(0.5,1)}, anchor=south,legend columns=4,legend cell align=left,font=\footnotesize},
    xtick=data,
    enlarge x limits=0.2,
    height=0.25\textwidth,
    width=0.48\textwidth,
    bar width=3mm,
    tick label style={font=\footnotesize},
    legend style={font=\footnotesize},
    ylabel={Trustworthy Rate},
    xlabel={countBF size (Megabyte)},
    symbolic x coords={9.1,18,36.19,70.92,142.2},
    ]
\addplot+[mygreen,mark=square*] table[x=memory,y=28error]{\memory};
\addplot [violet,thick,mark=*] table[x=memory,y=55error]{\memory};
\legend{28-mer,55-mer} 

\end{axis}
\end{tikzpicture}
\caption{\textbf{Comparison of the trustworthy rate among various KmerCo having different countBF size using 28-mers and 55-mer Balaenoptera dataset. Positive and close to zero is better.}} 
\label{figMT}
\end{figure} 

Figure \ref{figMT} expound on the performance of various KmerCo having different countBF sizes based on the trustworthy rate using the 28-mer and 55-mer Balaenoptera datasets. The figure adduces the issues of having a tiny-sized Bloom Filter. The insertion time decreases and the number of insertions per second increases with a decrease in KmerCo’s countBF size; however, small-sized countBF has more collisions. All KmerCo has a positive trustworthy rate. The KmerCo with 142.2 MB and 70.92 MB countBF has close to zero trustworthy rates. Others have close to zero trustworthy rates in the case of the 28-mer Balaenoptera dataset but in the case of the 55-mer Balaenoptera dataset, the trustworthy curve deviated upward indicating more erroneous K-mers considered as trustworthy K-mers. 

Overall, the KmerCo with 142.2 MB and 70.92 MB countBF showcased better performance based on insertion time, number of insertions per second, and trustworthy rate. However, KmerCo with 142.2 MB countBF is twice the size of KmerCo with 70.92 MB countBF. Along with better performance, the K-mer counting technique should maintain a small memory footprint. Thus, we considered the  KmerCo with 70.92 MB countBF for evaluating the performance of KmerCo using real datasets. 

\balance
